\documentclass[twocolumn]{AASTeX61}

\usepackage{mathrsfs}
\usepackage{xcolor}
\usepackage[T1]{fontenc}
\usepackage{fourier}
\usepackage{hyperref}
\usepackage[normalem]{ulem}
\usepackage{relsize}

\newcommand{\Msolar}{M$_\Sun$} 
 
\newcommand{\Mzams}{M$_{\text{ZAMS}}$}

\newcommand{\kspn}{KSP-SN-2016kf}
\newcommand{\iausn}{SN2017it}

\usepackage[english]{babel}															
\usepackage[protrusion=true,expansion=true]{microtype}	
\usepackage{amsmath,amsfonts,amsthm} 

\usepackage{subfigure}
\usepackage{natbib}
\shorttitle{Type II Supernova with Unusually High $^{56}$Ni Mass}
\shortauthors{Afsariadchi, Moon \& Drout et al.}

\begin{document}


\title{\kspn: a long-rising H-rich Type II Supernova with unusually high $^{56}$Ni mass discovered in the KMTNet Supernova Program}


\author{Niloufar Afsariardchi*}
\affil{Department {of Astronomy and Astrophysics, 50 St. George St., Toronto, ON M5S 3H4, Canada}, University of Toronto}

\author{Dae-Sik Moon}
\affil{Department {of Astronomy and Astrophysics, 50 St. George St., Toronto, ON M5S 3H4, Canada}, University of Toronto}

\author{Maria R. Drout}
\affil{Department {of Astronomy and Astrophysics, 50 St. George St., Toronto, ON M5S 3H4, Canada}, University of Toronto}
\affil{Hubble Fellow, Carnegie Observatories, 813 Santa Barbara Street, Pasadena, California, 91101 USA}
\affil{Dunlap Institute, University of Toronto, Toronto, ON M5S 3H4, Canada}

\author{Santiago Gonz\'alez-Gait\'an}
\affil{CENTRA, Instituto Superior T\'ecnico, Universidade de Lisboa, Av. Rovisco Pais 1, 1049-001 Lisboa, Portugal}

\author{Yuan Qi Ni}
\affil{Department {of Astronomy and Astrophysics, 50 St. George St., Toronto, ON M5S 3H4, Canada}, University of Toronto}

\author{Christopher D. Matzner}
\affil{Department {of Astronomy and Astrophysics, 50 St. George St., Toronto, ON M5S 3H4, Canada}, University of Toronto}

\author{Sang Chul Kim}
\affil{Korea Astronomy and Space Science Institute, 776 Daedeokdae-ro, Yusoeng-gu, Daejeon 34055, Republic of Korea}
\affil{Korea University of Science and Technology, Daejeon 34113, Republic of Korea}

\author{Youngdae Lee}
\affil{Korea Astronomy and Space Science Institute, 776 Daedeokdae-ro, Yusoeng-gu, Daejeon 34055, Republic of Korea}

\author{Hong Soo Park}
\affil{Korea Astronomy and Space Science Institute, 776 Daedeokdae-ro, Yusoeng-gu, Daejeon 34055, Republic of Korea}

\author{Avishay Gal-Yam}
\affil{Department of Particle Physics and Astrophysics, Weizmann Institute of Science, Rehovot, Israel, 76100}

\author{Giuliano Pignata}
\affil{Departamento de Ciencias Fisicas - Universidad Andres Bello, Avda. Republica 252, Santiago, 8320000 Chile}
\affil{Millennium Institute of Astrophysics (MAS), Nuncio Monsenor Sotero Sanz 100, Providencia, Santiago, Chile}

\author{Bon-Chul Koo}
\affil{Department of Physics and Astronomy, Seoul National University, Seoul 151-747, Republic of Korea}

\author{Stuart Ryder}
\affil{Department of Physics and Astronomy, Macquarie University, NSW 2109, Australia}

\author{Sang-Mok Cha}
\affil{Korea Astronomy and Space Science Institute, 776 Daedeokdae-ro, Yusoeng-gu, Daejeon 34055, Republic of Korea}
\affil{School of Space Research, Kyunghee University, 1732 Deogyeong-daero, Giheung-gu, Yongin-si, Gyeonggi-do, 17104, Republic of Korea}

\author{Yongseok Lee}
\affil{Korea Astronomy and Space Science Institute, 776 Daedeokdae-ro, Yusoeng-gu, Daejeon 34055, Republic of Korea}
\affil{School of Space Research, Kyunghee University, 1732 Deogyeong-daero, Giheung-gu, Yongin-si, Gyeonggi-do, 17104, Republic of Korea}

\email{*afsariardchi@astro.utoronto.ca}



\begin{abstract}

We present the discovery and the photometric and spectroscopic study of H-rich Type II supernova (SN) \kspn\  (\iausn) observed in the KMTNet Supernova Program in the outskirts of a small irregular galaxy at $z\simeq0.043$ within a day from the explosion. Our high-cadence, multi-color ($BVI$) light curves of the SN show that it has a very long rise time ($t_\text{rise}\simeq 20$~days in $V$ band), a moderately luminous peak ($M_V\simeq -$17.6~mag), a notably luminous and flat plateau ($M_V\simeq -$17.4~mag and decay slope $s\simeq0.53$~mag per 100 days), and an exceptionally bright radioactive tail. Using the color-dependent bolometric correction to the light curves, we estimate the $^{56}$Ni mass powering the observed radioactive tail to be $0.10\pm0.01$~M$_\sun$, making it a H-rich Type II SN with one of the largest $^{56}$Ni masses observed to date. The results of our hydrodynamic simulations of the light curves constrain the mass and radius of the progenitor at the explosion to be 
$\sim$15~M$_\sun$ (evolved from a star with an initial mass of $\sim$ 18.8~M$_\sun$)  
and $\sim1040$~R$_\sun$, respectively,
with the SN explosion energy of $\sim 1.3\times 10^{51}$~erg~s$^{-1}$.
The above-average mass of the \kspn\ progenitor, together with its low metallicity $\rm Z/Z_\Sun \simeq0.1-0.4$ obtained from spectroscopic analysis, is indicative of
a link between the explosion of high-mass red supergiants and their low-metallicity environment. 
The early part of the observed light curves shows the presence of excess emission above
what is predicted in model calculations, suggesting there is interaction between the ejecta
and circumstellar material.
We further discuss the implications of the high progenitor initial mass and 
low-metallicity environment of \kspn\ on our understanding of the origin of Type II SNe.

Online-only material: color figures
\end{abstract}


\section{Introduction}
\label{sec:intro}

Massive stars (\Mzams\ $\gtrsim$ 8 M$_\Sun$) end their lives with the collapse of their core, leading to supernova (SN) explosions in most cases. Type II SNe are those core-collapse SNe (CCSNe) whose progenitor retain part of their H envelope until the explosion. Some H-rich Type II SNe are marked with light curves featured with a distinctive post-peak plateau of $\sim$100 days. Following the often-undetected shock breakout (SBO) after the explosion, the envelope of the progenitor radiates shock-deposited energy \citep{Nakar2010,Rabinak2011}. After a few days, as the recombination front moves inside the ejecta,
the H recombination emission becomes the dominant emission powering the plateau \citep{Popov1993}.
This plateau emission is supplemented by contributions from the radioactive decay of $^{56}$Ni, 
depending on the $^{56}$Ni mixing level within the envelope \citep{Nakar2016}. Later, the energy released by the radioactive decay of $^{56}$Ni becomes the primary power source and the light curves enter 
the phase of the radioactive tail, followed by an optically thin nebular phase in a few months. Here, we avoid further classifying H-rich Type II SNe into traditional Type IIP and IIL sub-types due to increasing evidence that the transition between two classes is continuous \citep[e.g.,][and references therein]{Arcavi2017}.

Although red supergiants (RSGs) are largely considered to be the progenitor of H-rich Type II SNe, 
there still remains the critical uncertainty about the mass range of their progenitors. 
The RSGs identified in pre-explosion images of H-rich Type II SNe appear to have initial masses smaller than $\sim17$~M$_\Sun$, which is significantly lower than the upper limit of $\sim25$~M$_\Sun$ typically predicted in theories \citep{Smartt2015} as well as the observed RSGs in the Milky Way and Magellanic clouds \citep{Levesque2005,Levesque2006}. It has been argued that this discrepancy, called the `Red supergiant problem' \citep{Smartt2009}, may originate from unaccounted extinction in the circumstellar dust \citep{Walmswell2012} or inaccurate bolometric correction \citep{Davies2018}.
However, the origin of the lack of the observationally identified 
Type II progenitors in the mass range of 15--25 M$_\Sun$ remains still unknown.
There also exist theoretical uncertainties about the fate of massive stars, 
whose initial masses are in the range of $\gtrsim 17$~M$_\Sun$, and how they explode as SNe 
due largely to incomplete understanding of rotation, mixing, and mass-loss processes
as well as binary effect.
It may be possible that at least some RSGs in this mass range with a significant mass-loss rate
have their envelopes stripped, leading to more compact progenitors and, consequently, non-Type II
SNe such as Type Ib/c or IIb, although what fraction of the RSGs can have
such drastically substantial mass-loss rates is poorly understood  \citep{vanLoon2005,Smith2014}.
In addition, even the origins of the non-Type II -- i.e., Type IIb and Ib/c -- SNe 
are also somewhat uncertain
considering the lack of observational samples of progenitors of those SNe whose initial
masses are $\gtrsim 20$~M$_\Sun$ \citep[e.g.,][]{Drout2011,Lyman2016}
and the uncertainty of whether some of them originate from 
single stripped-envelope stars (e.g., Wolf-Rayet stars) or binary stars \citep{Smartt2009b}.
Interestingly, several recent studies suggest that 
massive RSGs in this mass range may rather produce a failed SN and implode to a black hole \citep{Oconnor2013,Sukhbold2016}
as potentially exemplified by the disappearance of a 25~M$_\Sun$ RSG in a nearby galaxy NGC 6946 \citep{Adams2017},
rendering the mapping between the progenitor masses and their final fates quite challenging.
We note, however, that metallicity may be an important factor determining how the massive RSGs explode as SNe. 
This is because most of the H-rich Type II SNe observed with a smaller, i.e., $\lesssim$~17 M$_\Sun$, progenitor mass
have been found in a host galaxy with a relatively high metallicity greater than 0.5 Z$_\Sun$,
while \citet{Anderson2018} reported the detection of a low-metallicity, i.e., 0.1~Z$_\Sun$, H-rich Type II SN 
whose initial progenitor mass is thought to be in the range of 17--25~M$_\Sun$.
This is suggestive that the RSGs whose initial masses are in the range of $\gtrsim 17$~M$_\Sun$
may prefer low-metallicity environments 
for them to explode as Type II SNe, and it is imperative to increase the observational
sample in order to support, or disapprove this tantalizing possibility.

Another important uncertainty in our understanding of the Type II SNe is the amount of $^{56}$Ni masses
produced in the explosion.
Statistical analyses based on the observed properties of the large number of recently-discovered Type II SNe 
have shown that they have a significant diversity in the observational characteristics such as rise time, peak brightness, decline rate, expansion velocity, and tail luminosity  \citep{Arcavi2012,Anderson2014, Sanders2015, Gonzalez2015,Rubin2016,Valenti2016}. 
The mass of $^{56}$Ni produced in Type II SN explosions
obtained from the analysis of their tail luminosities 
is considerably larger than what is predicted by neutrino-driven CCSN simulations \citep{Ugliano2012,Pejcha2015a,Sukhbold2016}, 
suggesting that our understanding of how Type II SNe explode is incomplete.
The identification and detailed observational studies of Type II SNe with a large 
$^{56}$Ni mass, therefore, can provide valuable insights into their 
progenitors and explosion mechanisms.

It is worthwhile to note that the observational studies of SNe, including H-rich Type II, often rely on insufficiently-sampled light curves without early coverage and color information. However, early rise time light curves contain vital information on key progenitor parameters, especially progenitor radius \citep[e.g.,][]{Rubin2016,Rubin2017}. Furthermore, the small number of SNe observed with early light curves make it difficult to
investigate important processes involved in the explosions, including
potential interactions with a binary companion \citep{kasen2010},
aspherical behaviours \citep{Afsariardchi2018}
as well as mass-loss history of progenitors prior to the explosion \citep[e.g.,][and references therein]{Yaron2017}.




Indeed, the growing number of early SN detections have shown strong indications for
the presence of the circumstellar material (CSM) at the vicinity of their progenitors, including excess emission in the early light curve compared to the predictions of SBO cooling emission model \citep{Morozova2018,Forster2018} as well as  ``flash-ionized'' narrow H or He emission lines in the early spectra that disappear within hours or days following the explosion \citep{Galyam2014, Khazov2016, Yaron2017,Bullivant2018, Hosseinzadeh2018}. These facts suggest that the progenitors of these SNe may have undergone a short period of enhanced pre-SN mass loss or outbursts in the final years leading to core collapse. The nature of this dense CSM is still unclear, but one promising mechanism that can lead to its formation is pre-SN wave heating outbursts \citep{Quataert2012,Fuller2017, Ro2017, Fuller2018}. High-cadence early observations of SNe are crucial for constraining the structure of such CSM and shining light on the types of SNe experiencing enhanced pre-SN mass loss.

In this paper, we present the discovery and observational studies, supplemented by numerical simulation work, of a H-rich Type II SN \kspn\ that we detected at a very early epoch, likely within $\sim$1 day from the explosion, using the high-cadence, multi-color data obtained
in the KMTNet (Korea Microlensing Telescope Network) Supernova Program \citep{Moon2016}. Our multi-color ($BVI$) coverage from the early epoch allows us to model the early light curves and precisely measure the explosion epoch and the rise time of the light curves, leading to important insights into its progenitor, especially the progenitor radius and potential CSM characteristics.

This paper is organized as follows. In $\S$\ref{sec:phot} we describe the discovery and photometric and spectroscopic observations of \kspn, which is followed by light curve analysis and the estimation of $^{56}$Ni mass in $\S$\ref{sub:LC}.  $\S$\ref{sec:spec} and $\S$\ref{sub:host} provides the spectroscopic analysis of the SN and its host, respectively, whereas  $\S$\ref{sec:prog} does the hydrodynamic simulations for the progenitor and potential interaction with the CSM. We discuss the implications of our results for understanding the Type II SNe in $\S$\ref{sec:diss}, and conclude in $\S$\ref{sec:sum}.

\section{Observations and Discovery}
\label{sec:phot}

\subsection{Photometry and Discovery}
\label{sub:phot}
\kspn\ (\iausn\footnote{https://wis-tns.weizmann.ac.il/object/2017it.
Note that although the first detection of \kspn\ 
was made in 2016 by the KMTNet as we report in this paper, 
the official registration of the source in the Transient Name Server 
as an astronomical transient was made by Gaia Alerts team based on its detection of the source 
on January 5, 2017.})
was first detected on December 24, 2016 as part of the KMTNet Supernova Program \citep[KSP;][]{Moon2016}.
The KMTNet is a network of three 1.6~m telescopes located in Chile, South Africa, and Australia, providing 24-hour continuous  sky coverage. Each telescope of the KMTNet is equipped with a wide field CCD camera covering $2 \degr\times 2 \degr$ field at $0 \farcs$4 pixel sampling \citep{Kim2016}. The KSP conducts high-cadence $BVI$ monitoring of a sample of fields focusing on studying early supernovae and  rapidly evolving transients \citep[e.g.,][]{heet16,antoet17,parket17,leeet18,broet18}.
Between 2016 October and 2017 May, we obtained about 600 images per each $BVI$ band, with a mean cadence of roughly 9 hours per band, for a field around the nearby galaxy NGC~2188.
All the images were obtained with 60-s exposure time, reaching $\sim$ 20.5--22.5 limiting magnitudes\footnote{All magnitudes are reported in Vega system.} depending on the filter and observational conditions.

\begin{figure*}[th]
\centering
\begin{center}
\includegraphics[scale=0.5]{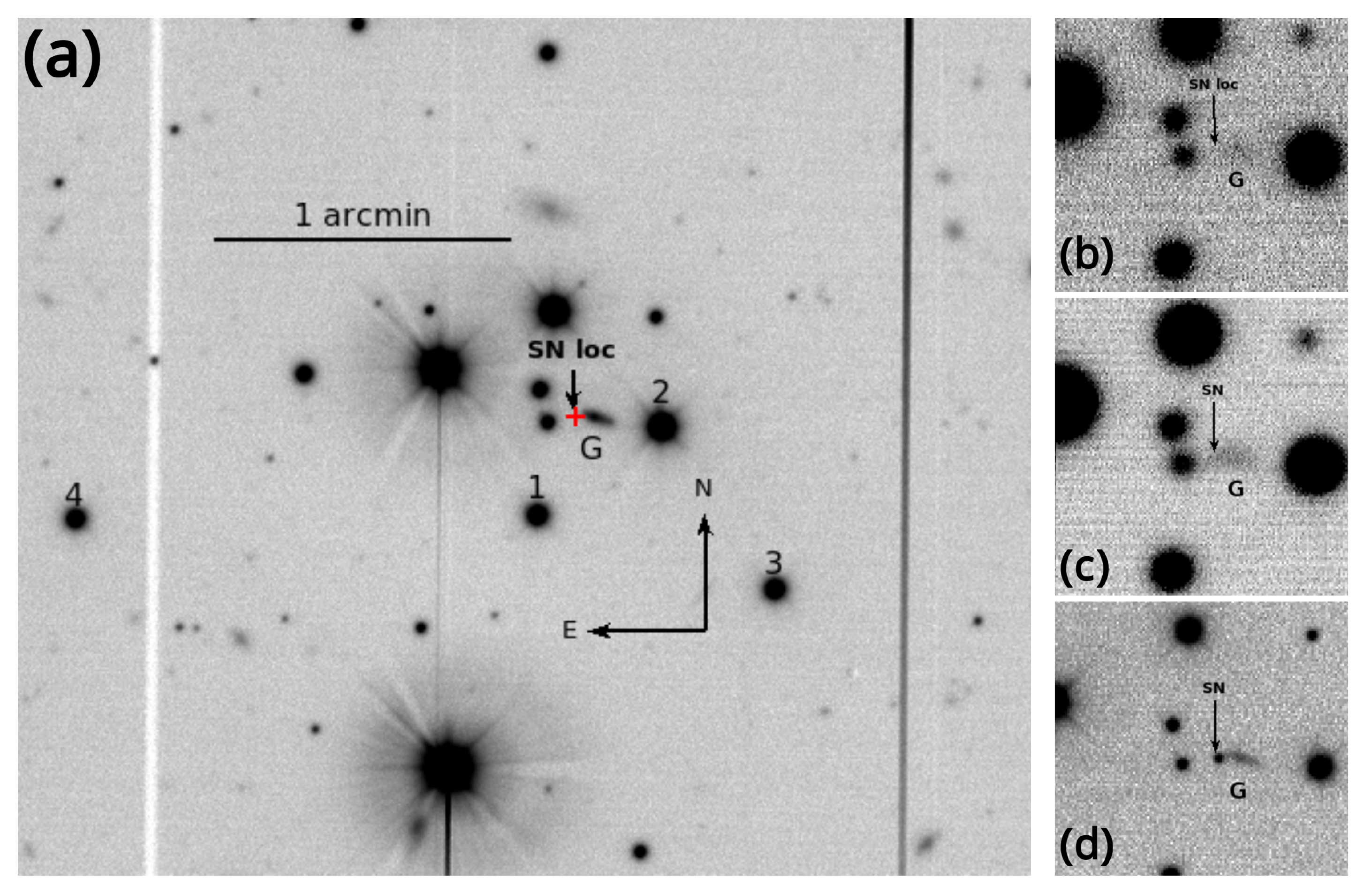}
\end{center}
\centering
\caption{ $B$-band images of \kspn\ field. (a) A deep stacked image of the field made from 114 high quality images from 2016 October until 2016 December before the SN explosion. The host galaxy of \kspn\ is indicated by letter``G'' and the SN position is identified by a red cross. The numbered stars are a few of the standard stars used for photometric calibration. On the right panel, the evolution of \kspn\ is shown in $B$-band at different epochs including (b) the last  non-detection at MJD=57745.488, (c) the first detection at MJD=57746.867, and (d)  around maximum brightness at MJD=57759.146.}
\label{fig:host}
\end{figure*}

Figure~\ref{fig:host} shows $B$-band images of a field containing \kspn, including (a) a deep stack image of a $\sim 4\arcmin \times 4\arcmin$ field  around the source, (b) the last  image before the detection of \kspn\ at MJD=57745.488, (c) the first detection image of \kspn\ at MJD=57746.867 and (d) around peak brightness at  MJD=57759.146.
As in Figure~\ref{fig:host}(b), \kspn\ was first captured in the $B$-band with magnitude of $\sim20.4$ at (R.A., decl.) (J2000) =(${\rm 06^h10^m37.156^s}$,$-34\degr08\arcmin28\farcs3$) in the outskirts of an elongated faint irregular galaxy $\sim$4\arcsec\ (or $\sim$3.6 kpc at the angular diameter distance of 181.5~Mpc, see below $\S$\ref{sub:red} for the measurement of distance to the galaxy) away in the western direction. See $\S$\ref{sub:host} for the identification of the galaxy as the host galaxy of \kspn\ based on spectroscopic information. 


The KSP real-time data processing pipeline first performs the bias subtraction, cross talk removal, and flat-fielding of the science images. Next, the astrometric solution is obtained by \texttt{SCAMP}\footnote{http://www.astromatic.net/software/scamp} \citep{Bertin2006} using $\sim$10000 unsaturated stars which have a counterpart in the second Hubble guide star catalogue \citep{Lasker2008}, resulting in precision of $\sim$0\farcs12. 

For the \kspn, we subtracted from the science images reference images made from pre-SN images of the field using the \texttt{HOTPANTS}\footnote{http://www.astro.washington.edu/users/becker/v2.0/hotpants.html}  program \citep{Becker2015}. Since the SN is close to its host galaxy, image subtraction is crucial for reducing the effect of the host galaxy on the SN photometry. By adjusting \texttt{HOTPANTS} parameters, we ensured that the subtraction is done efficiently and does not introduce bias in the subtracted science images. The photometry was then carried out on subtracted science images using our custom-developed software that performs PSF fitting and obtains the photometric solution based on $\sim$35 AAVSO All-Sky Survey nearby standard stars \footnote{https://www.aavso.org/apass}. The photometry of \kspn\ before applying any corrections is provided in Table \ref{tab:photometry}.



\begin{deluxetable}{cccc}
\tabletypesize{\footnotesize}
\tablecolumns{4} 
\tablewidth{0.99\textwidth}
 \tablecaption{\kspn\ optical photometry \label{tab:photometry}}
 \tablehead{
 \colhead{Time [MJD]} & \colhead{Filter} & \colhead{Magnitude$^1$ [mag]} & \colhead{Error [mag]} 
 } 
\startdata 
57745.488 & $B$ & >21.01 & - \\
57745.489 & $V$ & >21.04 & - \\
57745.490 & $I$ & >20.62 & - \\
57746.867 & $B$ & 20.56 & 0.14  \\
57746.868 & $V$ & 20.41 & 0.15  \\
57746.869 & $I$ & 20.36 & 0.23  \\
57747.482 & $B$ & 19.98 & 0.09  \\
57747.483 & $V$ & 19.96 & 0.09  \\
57747.485 & $I$ & 20.06 & 0.18  \\
57747.867 & $B$ & 19.74 & 0.04  \\
57747.869 & $V$ & 19.64 & 0.05  \\
57747.870 & $I$ & 19.68 & 0.08  \\
\enddata
\tablenotetext{1}{Magnitudes are in Vega system and not corrected for extinction. }
 \tablecomments{Table\ref{tab:photometry} is published in its entirety in the electronic edition. A
portion is shown here for guidance regarding its formatting.}
\end{deluxetable} 

\subsection{Spectroscopy}
\label{sub:introsprec}
We obtained a spectrum of \kspn\ on March 27, 2017, which is 93 days after the first detection, using the WFCCD spectrograph mounted on the 2.5-m du Pont telescope at Las Campanas Observatory.  The spectrum was acquired with the low resolution ``blue'' grism and a 1\farcs 65 slit aligned with the parallactic angle, providing a wavelength coverage of $\sim$3500$-$9000~\AA. 

Bias and flat field correction, sky subtraction, spectral extraction, and wavelength calibration were performed using standard tasks in IRAF\footnote{IRAF is distributed by the National Optical Astronomy Observatory, which is operated by the Association for Research in Astronomy, Inc.\, under cooperative agreement with the National Science Foundation.} \citep{Tody1993}. In addition to the spectrum of \kspn\, we also extracted a spectrum of the host galaxy from the same data.  Flux calibration and telluric correction of both spectra were performed using a set of custom IDL scripts (see, e.g., \citealt{Matheson2008,Blondin2012}). The observation of standard stars for flux calibration was made after the SN observation on the same night. The observed spectra and the results of our spectroscopic analysis are presented in $\S$\ref{sec:spec}.




\subsection{Redshifts, Reddening Correction and $K$-Corrections}
\label{sub:red}
We measure the redshift of the host galaxy of \kspn\ to be 0.043 $\pm$ 0.002 using the  H$\alpha$ line in the host galaxy spectrum. 
The value is consistent with the redshift measured with the SN spectrum, 
confirming that the irregular galaxy G (Figure~\ref{fig:host}) is the host galaxy of \kspn\ 
(see $\S$~4 and 5 for the details of the spectroscopic analysis and the redshift measurements). 
In this paper we adopt the standard $\rm{\Lambda CDM}$ cosmology with Hubble constant $H_0=67.4$~km~s$^{-1}$~Mpc$^{-1}$, matter density parameter $\Omega_M=0.315$, and vacuum density parameter $\Omega_\Lambda=0.685$ \citep{PlanckCollaboration2018}. 
The luminosity distance and the distance modulus of the host galaxy based on these parameters are 
197.4~Mpc and 36.48~mag, respectively. 
Table~\ref{tab:gen_params} contains a list of the key parameters of \kspn\ obtained in our analysis.

Since \kspn\ is located in the outskirts of the host galaxy, 
it is highly unlikely that there exists any substantial host galaxy extinction.
We, therefore, only consider the extinction from the Milky Way in our reddening correction. 
This is supported by the absence of the
\ion{Na}{1}D $\lambda \lambda$5890, 5896 doublet absorption feature,
which is known to be indicative of host galaxy extinction \citep[e.g.][]{Poznanski2012},
in our SN spectrum (see \S\ref{sec:spec}).
Our extinction measurement using the observed Balmer Decrement also shows
that the host galaxy extinction is negligible (see $\S$\ref{sub:host} for the details).
We obtain the Galactic extinction $E(B-V) \simeq$0.029~mag toward the location of \kspn\ using
the extinction model of \citet{Schlafly2011} and conduct a reddening correction 
of the extinction assuming $A_V/E(B-V)=3.1$. 

In addition to the reddening correction, 
we also carry out $K$-corrections to obtain the final photometric solution 
using the default $(1+z)$ $K$-Correction factor \citep{Oke1968}. 
Note that we ignore the effect of the SED shape on the $K$-corrections 
as our photometric solution based on the SN spectrum indicates the effect 
is negligible at the redshift of \kspn. 
We also ignore the time dilation effect given its low redshift.

\begin{figure*}[!t]
\centering
\includegraphics[scale=0.75]{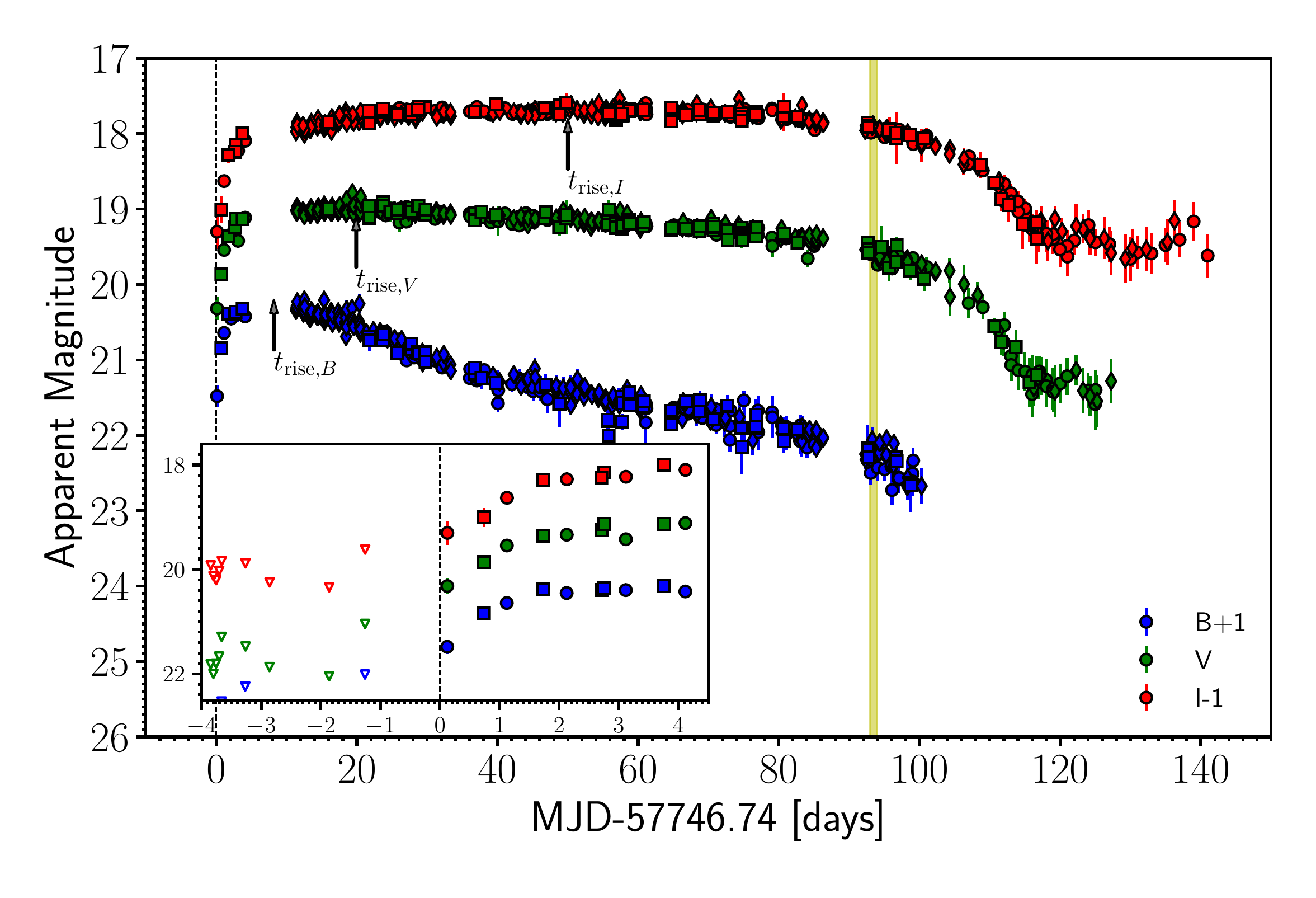}
\caption{Light curves of \kspn\ in $BVI$ bands, vertically shifted for readability. The x-axis represents time since our estimated $t_\text{SBO}=57746.74$. The filled circle, diamond, and square markers indicate the observations made by South African, Chilean, and Australian telescopes, respectively. The open inverted triangles in the zoomed-in version of the early light curve mark non detection upper limits. The yellow line represents the spectroscopy epoch and the dashed line is our estimated SBO time. The rise times in $BVI$ are marked by vertical arrows. Note that the light curves are corrected for Milky Way extinction $E(B-V)=0.029$~mag \citep[from][]{Schlafly2011}. }
\label{fig:lc}
\end{figure*}

\begin{figure}[t!]
\centering
\includegraphics[trim=0.8cm 0.5cm  0.5cm 0.5cm, clip=true, scale=0.53]{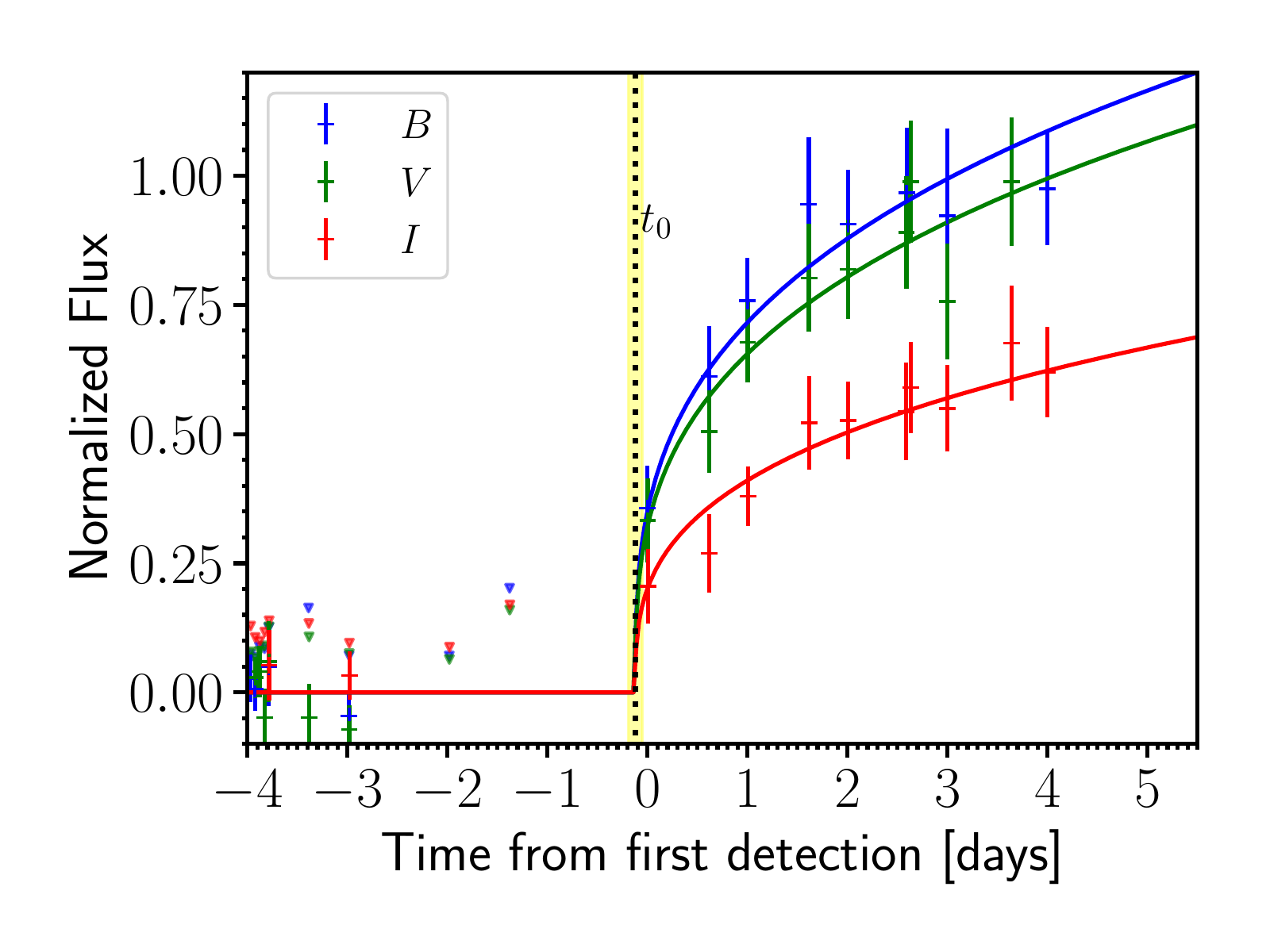}
\caption{Estimated epoch of SBO. We fit a power law to the fluxes normalized to their corresponding peak values in $B$ (blue curve), $V$ (green curve), and $I$ (red curve) bands. We only consider the epochs with fluxes within 5 days of the first detection including upper limits of non-detection epochs for fitting the epoch of first light. We mark the best fitted epoch of first light $t_0\simeq-0.12$~days with dotted vertical line, while its 1-$\sigma$ confidence level of $0.07$~days is shown with yellow strip. The non-detection limiting fluxes are shown with inverted triangles. }
\label{fig:t0}
\end{figure}

\begin{figure*}[t!]
\centering
\includegraphics[trim=1cm 1cm  0.5cm 0.5cm, clip=true, scale=0.9]{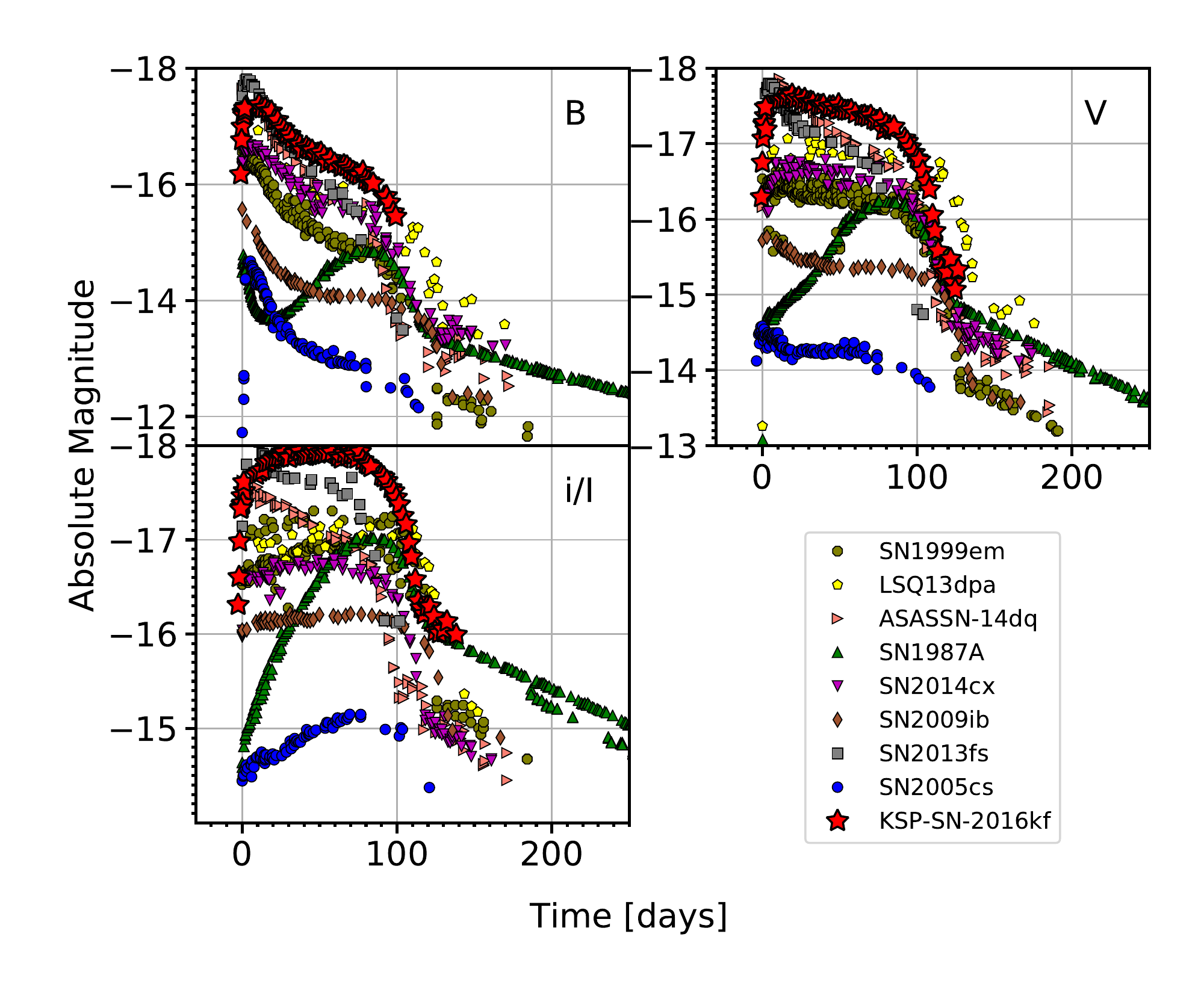}
\caption{The light curve of \kspn\ is compared against other well-observed or similar Type II SNe: SN1999em \citep{Elmhamdi2003} LSQ13dpa \citep{Valenti2016}, ASASSN-14dq \citep{Valenti2016}, SN1987A \citep{Hamuy1988}, SN2014cx \citep{Huang2016}, SN2009ib \citep{Takats2015}, SN2013fs \citep{Yaron2017}, and SN2005cs \citep{Pastorello2009}. Here, the light curves of \kspn\ are binned into 1 day intervals to reduce the variance of the data points. The x-axis is time since $B$-band peak magnitude for each SNe (except for SN1987A which is presented relative to the first $B$-band detection).  The shape of the light curves of \kspn\ resembles than of SN2014cx, SN1999em, and LSQ13dpa, but is notably more luminous; in particular, \kspn\ has the most luminous plateau (measured at 50 days) and tail luminosity in comparison with other Type II SNe shown here.}
\label{fig:comp}
\end{figure*}

\begin{deluxetable}{lcc}
\tabletypesize{\footnotesize}
\tablecolumns{2} 
\tablewidth{0.99\textwidth}
 \tablecaption{\kspn\ parameters \label{tab:gen_params}}
 \tablehead{
 \colhead{Parameter} & \colhead{Value} 
 } 
\startdata 
Redshift, $z$ &  0.043 $\pm$ 0.002  &
\\
Reddening $^{1}$, $E(B-V)$ [mag] & 0.029 $\pm$ 0.001 &
\\
Shock breakout time, $t_\text{SBO}$ [MJD] & 57746.74 $\pm$ 0.07 & 
 \\
Offset from host galaxy, $r$ [kpc] &   3.4 $\pm$ 0.2 & 
\\
Luminosity distance, $d_L$ [Mpc] &    197.4 $\pm$ 10.9 & 
\\
Angular diameter distance, $d_A$ [Mpc] &   181.5 $\pm$ 9.3 & 
\\
Distance modulus, $\mu$ [mag] & 36.48 $\pm$ 0.12 &
\\
$^{56}$Ni mass , $M_{\text{Ni}} [M_\Sun]$& 0.10 $\pm$ 0.01 &
\enddata
\tablenotetext{1}{Galactic dust extinction from \cite{Schlafly2011} }
\end{deluxetable}

\section{Light Curve Analysis}
\label{sub:LC}

\begin{deluxetable*}{lccccccc}
\tabletypesize{\footnotesize}
\tablecolumns{4} 
\tablewidth{0.99\textwidth}
 \tablecaption{ Light Curve Parameters of \kspn\
 \label{tab:lc}}
 \tablehead{
 \colhead{ } & \colhead{$B$ } & \colhead{$V$ } & \colhead{$I$ }
 } 
\startdata 
  Rise Time, $t_\text{rise}$[days] & $8.2 \pm$ 3.6 & 19.9 $\pm$ 1.2 & 50.0 $\pm$ 1.2  & 
  \\
 Bazin$^1$ Rise Time, $t_\text{baz}$[days] & 7.93 $\pm$ 4.51 & 10.78 $ \pm$ 6.62 & 31.22 $\pm$ 10.72  & 
  \\
Peak Absolute Magnitude, $M_\text{peak}$ [mag] & -17.40 $\pm$ 0.16 & -17.62 $\pm$ 0.14  & -17.92 $\pm$ 0.13 & 
 \\ 
Absolute Magnitude at 50 days, $M_{50}$ [mag] & -16.34 $\pm$ 0.15 & -17.47 $\pm$ 0.14 & -17.92 $\pm$ 0.13  & 
\\
Decay Rate, $s$ [mag/100 days]  & 1.37 $\pm$ 0.07 & 0.53 $\pm$ 0.04 & 0.02 $\pm$ 0.01 &
\\
Tail Absolute Magnitude, $M_\text{tail}$ [mag] & -- & -15.40 $\pm$ 0.15 & -16.30 $\pm$ 0.15 &
\enddata

 \tablecomments{The quantities are measured with respect to the epoch of first light at 57746.74 MJD.}
 \tablenotetext{1}{Obtained by fitting a functional form proposed in \cite{Bazin2009}}
\end{deluxetable*}

\subsection{Light Curve Evolution}
\label{sub:earlyLC}

As explained above, \kspn\ was first detected on December 24, 2016,  MJD = 57746.867 in the $B$ band with $20.56 \pm 0.14$~mag followed by detections in the $V$ and $I$ bands, 2 and 4 minutes later, respectively, with $20.41 \pm 0.15~(V)$ and $20.36 \pm 0.23~(I)$~mag. The last $BVI$ non-detection images were obtained $\sim$33 hours before the first detections with limiting magnitudes of $\sim$21.01~$(B)$, $\sim$21.04~$(V)$, $\sim$20.62~$(I)$~mag at the 3-$\sigma$ confidence level. 

Figure \ref{fig:lc} provides the entire light curves that we obtained for \kspn\ with KMTNet, spanning over 140 days. As can be seen in Figure \ref{fig:lc}, the $V$-band light curve of \kspn\ rises for $\sim$20~days (see below for the precise rise time measurement) before entering a plateau phase of $\sim$105~days during which the light curve slowly declines. Over the interval of $\sim$90--110 days from the SBO, the light curves dim for $\sim$1.5~mag and transition into a slowly-decaying tail. The described light curve evolution and, in particular, the prominent plateau phase of \kspn, together with the timescales of its distinctive phases of evolution, identify this transient as a H-rich Type II SN
\citep[Type IIP SN,][]{Arcavi2012}. Note that the SN nature is also confirmed by the spectroscopic information presented in $\S$\ref{sec:spec}. 


In order to obtain the epoch of first light of \kspn, which is the SBO epoch for CCSNe \citep{Matzner1999}, we fit the early part of the light curves by the following equation
\begin{eqnarray}
N_\lambda &=& \begin{cases}
C_\lambda (t-t_0)^n  &\hspace{1cm} t>t_0\\
0 &\hspace{1cm}  t\leq t_0
\end{cases},
\end{eqnarray}
where $N_\lambda$ is normalized flux by its peak value in each band, $t$ time measured from the epoch of the first detection, $t_0$ the epoch of SBO relative to the epoch of the first detection, $n$ power index, and $C_\lambda$ fitting coefficient. We only consider the epochs within 5~days from the first detection as well as upper limits for the epochs of non-detections in the fitting procedure. We fit the $BVI$ light curves of \kspn\ simultaneously and find the following best-fit parameters: $t_0= -0.12 \pm 0.07$ days, $(C_B, C_V, C_I)= (0.69 \pm 0.04, 0.63 \pm 0.03, 0.39 \pm 0.02)$, and $n=0.32 \pm 0.04$. Figure \ref{fig:t0} compares the best fit power laws with the observed $BVI$ early light curves, where the SBO epoch $t_0=-0.12$ days is marked with vertical dotted line. In calendar days, this SBO epoch corresponds to  $t_\text{SBO}=57746.74 \pm 0.07$ MJD (i.e., December 24, 2016).

For measuring the rise time of \kspn, we use two methods: 1) $t_{\text{rise}}$ is the time from $t_\text{SBO}$ to the maximum light  $t_{\text{peak}}$ found by fitting a low order polynomial to the observed light curves; 2) $t_{\text{baz}}$ is defined as the time from $t_\text{SBO}$ to the maximum of the \cite{Bazin2009} exponential functional form fitted to the observed light curves. Table~\ref{tab:lc} summarizes the best-fit $t_\text{rise}$ and $t_{\text{baz}}$. We found that $t_{\text{baz}}$ is smaller than $t_{\text{rise}}$ for \kspn. The difference between the two times is notable for $I$ band, for which the early light curve exhibits two distinct slopes over 50 days: an initial steep rise during the first $\sim$20 days and a slow rise to the peak during which the magnitude changes only $\sim$0.3 mag over $\sim$30 days. Note that $t_{\text{baz}}$ tends to represent the time scale of the former slope \citep{Gonzalez2015} and therefore it is particularly shorter than $t_{\text{rise}}$ in the $I$ band. In Table~\ref{tab:lc}, the high uncertainty in the $B$-band $t_{\text{rise}}$ is due to the lack of data between $\sim 4$ and $\sim 10$ days post-discovery.








After the rise, the light curves reach the peak magnitude $M_\text{peak}\simeq -17.40 \ (B), -17.62 \ (V)$, and  $-17.92 \ (I)$~mag, followed by the plateau phase. During the plateau phase, the light curves decay slowly, especially in the $V$ and $I$ bands. We measure the plateau length of $\sim$105 days by fitting a Fermi-Dirac function to the transition part of the light curves \citep{Valenti2016}. Table~\ref{tab:lc} contains the $BVI$ decay rates of the light curves during the plateau phase. Between the epoch of $\sim90$ and $\sim 118$ days, the light curves rapidly decline by more than 2~mag in the $V$ and $I$, after which  they settle into a radioactive tail, lasting for an extended period of time with a slow, gradual dimming.

\subsection{Light Curve Comparison}
\label{sub:lccomp}
Figure~\ref{fig:comp} compares the light curves of \kspn\ with those of eight other Type II SNe in $BVI$ band -- all of them are of H-rich Type II, except for SN1987A classified as Type II-Peculiar. The epochs are relative to $B$-band peak brightness for each SN; except for SN1987A, which are relative to the first $B$-band detection due to its double-peaked light curve shape. We can identify in Figure \ref{fig:comp} that there exists a significant diversity among Type II SNe in their luminosities and decay rates. \kspn\ has highest luminosity of the other SNe shown: in particular, both its plateau and
tail luminosities are higher. Furthermore, the decay rate of \kspn\ is among the smallest which makes the plateau look nearly flat in the $I$ band. In terms of the shape of the light curves, \kspn\ resembles those of SN2014cx, SN1999em, and LSQ13dpa but is notably more luminous. The $I$-band light curve of \kspn\ has a peculiar bell shape, which has been also observed for SN2014cx.


Comparing the rise time of \kspn\ to the rise time distribution of \cite{Gonzalez2015}, we find that the $V$- and $I$-Band rise times of \kspn\ are, respectively, more than 1-$\sigma$ and 2-$\sigma$ above the median of $t_{\text{rise}}$---i.e., 11.5~($V$) and 19.3~($I$) days. Likewise, $t_\text{baz}$ of \kspn\ in $I$ band is more than 3-$\sigma$ above than the median of Type II SNe, while in $B$ and $V$ bands, the $t_\text{baz}$ is within 1-$\sigma$ of the median. Based on this high value of $t_{\text{rise}}$, \kspn\ can be considered a ``long riser'' Type II SN in the definition of \cite{taddia2016}. 

Additionally,  the plateau of \kspn\ at 50 days is among the most luminous ones observed with absolute magnitude $M_V \simeq -17.47$~mag, while its peak magnitude is only moderately bright at $M_V \simeq -17.62$~mag. This highlights that the decay rate of \kspn\ is relatively small. The decay rates  of Type II SNe are often characterized by two slopes: an initial steep slope that immediately follows the light curve peak and a smaller decay rate that precedes the end of the plateau \citep{Anderson2014}. However, as \cite{Valenti2016} highlights, not all H-rich Type II SNe exhibit distinct slopes during the plateau phase. Indeed, \kspn---similar to SN2014cx and prototypical SN1999em---shows nearly constant decay at the rate $s\simeq0.53$~mag per 100 days in $V$ band in the interval 30--70 days.

The decay rate and peak magnitude of Type II SNe are strongly correlated \citep{Sanders2015}. We test whether \kspn\ also follows this correlation by setting $s=0.53$ mag per 100 days in the peak magnitude--decay rate relation of \cite{Anderson2014}. This gives the expected $V$-band peak magnitude of $\sim-16.58$~mag which is $\sim1$ magnitude fainter than the observed $V$-band $M_\text{peak}\simeq-17.62$~mag presented in Table \ref{tab:lc}, indicating that \kspn\ is not compatible with this correlation.


\subsection{Color Analysis}

\begin{figure}[t!]
\centering
\includegraphics[trim=0.5cm 1.3cm  0.5cm 0.7cm, clip=true, scale=0.56]{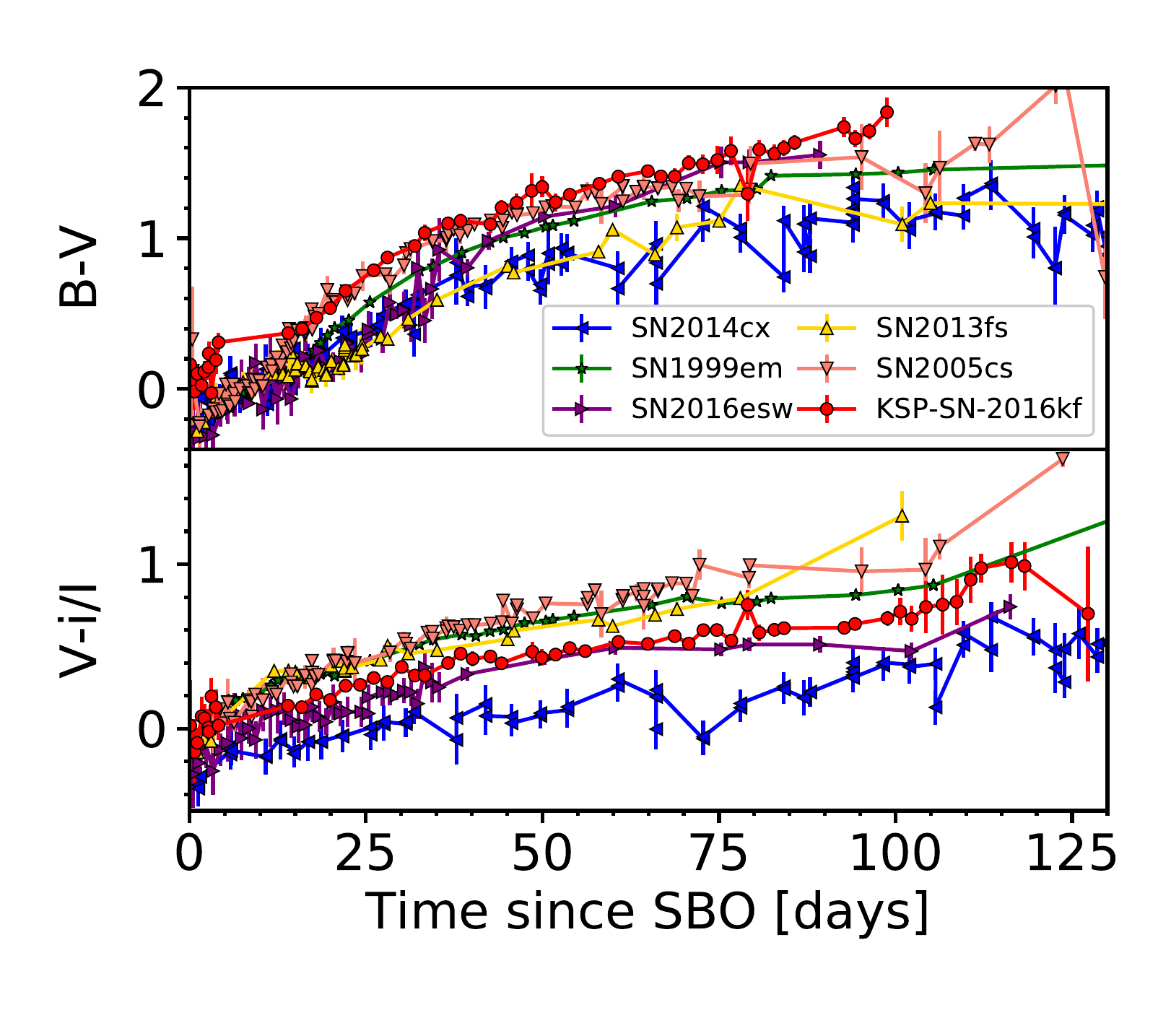}
\includegraphics[trim=0.8cm 0.8cm  0cm 0.7cm, clip=true, scale=0.57]{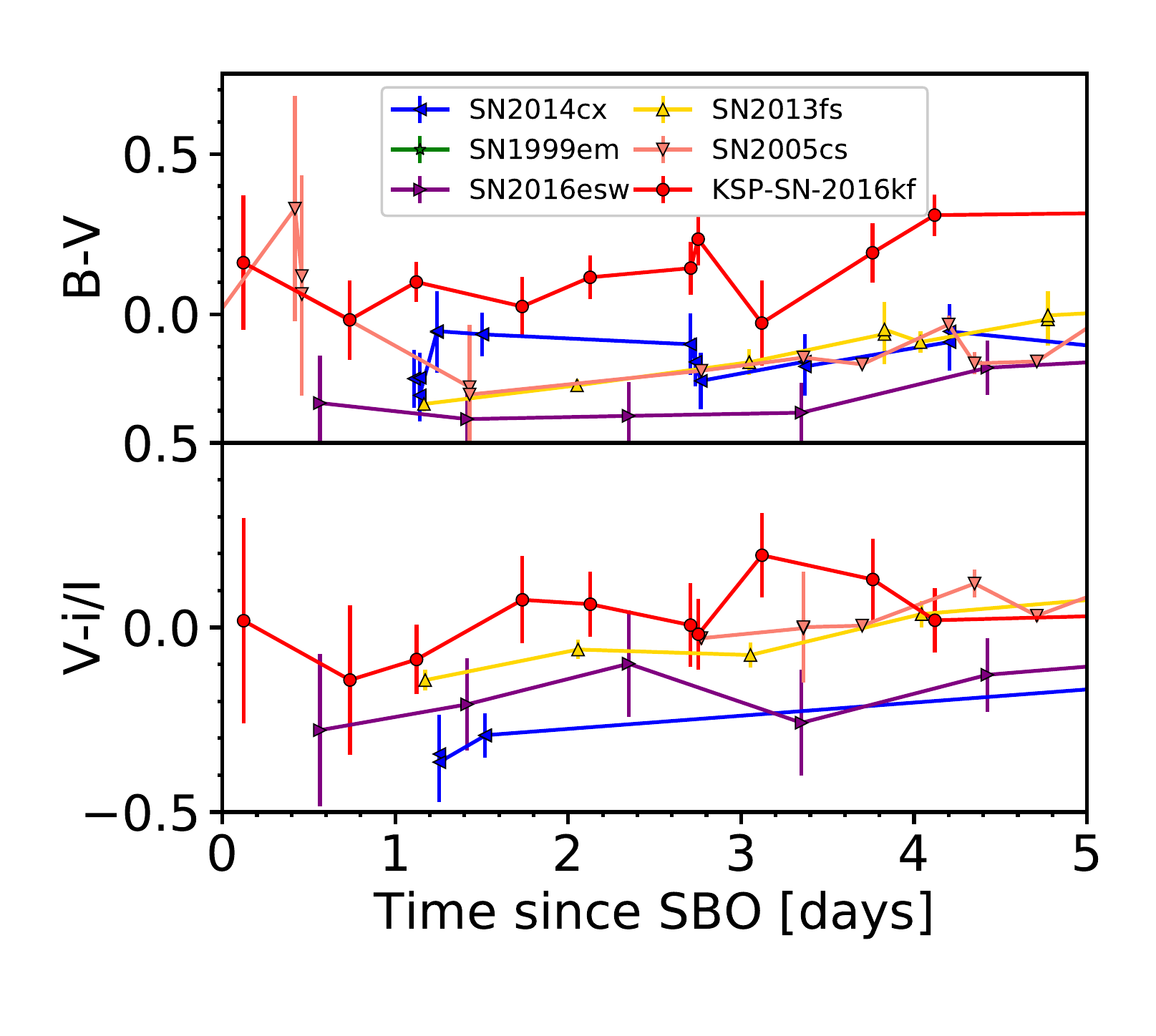}
\caption{The color index of \kspn\ along with other well-observed or similar Type II SNe: SN2014cx \citep{Huang2016}, SN1999em \citep{Elmhamdi2003}, SN2016esw \citep{deJaeger2018}, SN2013fs \citep{Yaron2017}, and SN2005cs \citep{Pastorello2009}. \emph{Top:} the $B-V$ and $V-i/I$ of SNe are shown for less than 120 days since SBO. \emph{Bottom:} the same as above but zoomed-in on the first week evolution.}
\label{fig:cindex}
\end{figure}

Figure~\ref{fig:cindex} (top panel) compares the $B-V$ and $V-i/I$ color evolution of \kspn\ with those of a sample of other H-rich Type II SNe for the entire 130 days evolution, while Figure~\ref{fig:cindex} (bottom panel) presents the same for the first 5 days of the evolution. Overall, the color evolution of \kspn\ is similar to other shown SNe confirming its H-rich Type II nature. As in the Figure, the colors of these SNe redden during the first $\sim$ 50 days in an almost linear manner, after which the color change rates become small before they enter the phase of the radioactive tail. While the color evolution of \kspn\ is broadly similar to that of other prototypical H-rich Type II,  \kspn\ is redder than the other SNe, especially in $B-V$ and during the first 5 days. The redder color of \kspn\ does not appear to follow the  correlations between the peak absolute magnitude and color in \cite{deJaeger2018b}. Based on this correlation, it is expected that fainter Type II SNe (in terms of peak absolute magnitude) will be redder in color, although it is worth noting that the dispersion on this correlation is quite large and as pointed by \cite{deJaeger2018b} the intrinsic differences of the SN progenitors, including progenitor radius and the CSM characteristics, are the contributing factors in the dispersion of the color. Note that the color of \kspn's radioactive tail is only shown in the $V-I$ plot due to lack of $t\gtrsim100$ days data in $B$ band. During the tail, $V-I$ rapidly reddens by $\sim0.3$ mag over $\sim$10 days. A similar behaviour has been previously observed for other Type IIP SNe \citep[e.g., SN2005cs;][]{Pastorello2009}.


\begin{figure}[t!]
\centering
\includegraphics[trim=0.1cm 0.1cm  0cm 0.5cm, clip=true, scale=0.57]{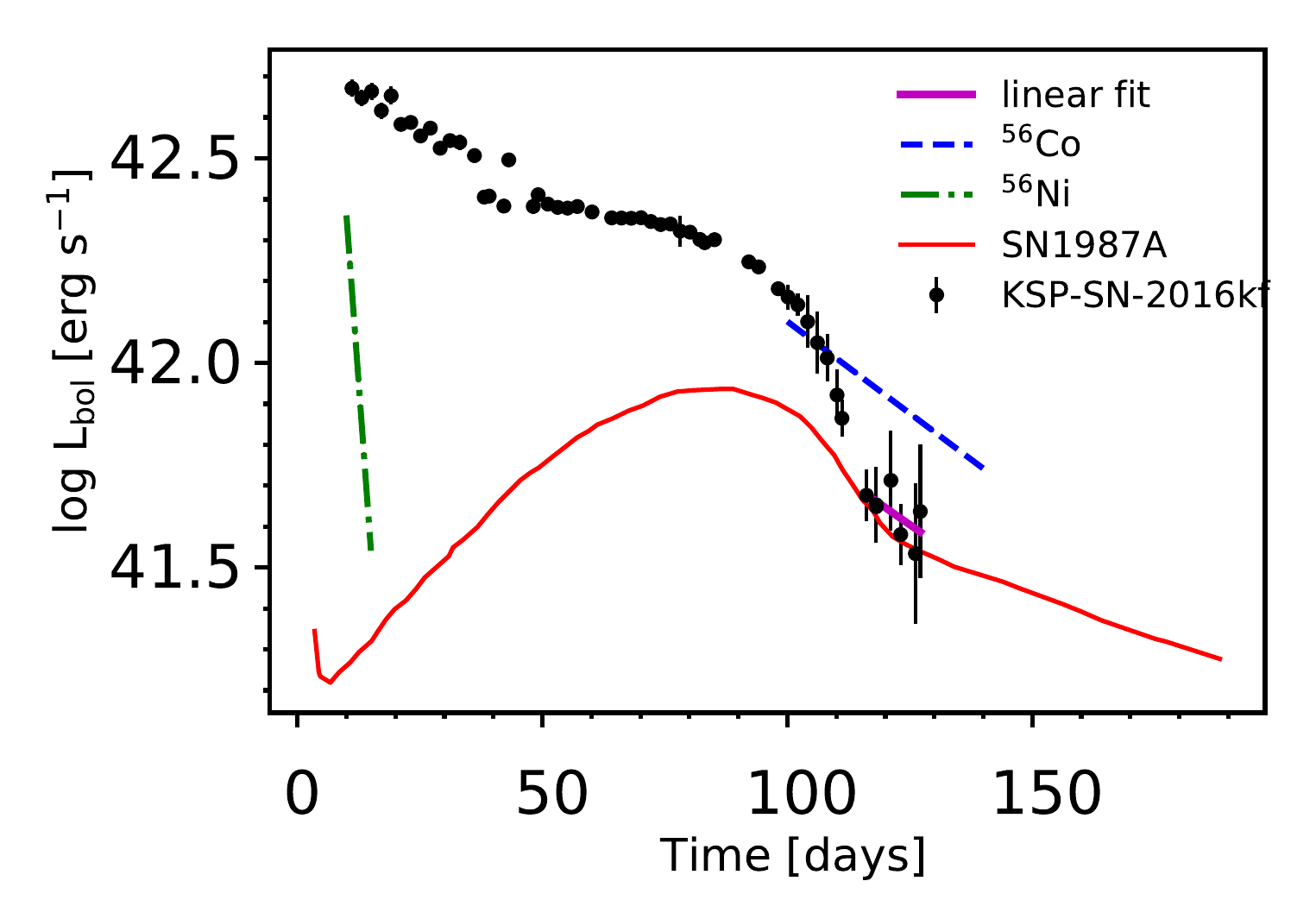}
\caption{Bolometric light curve of \kspn\ (black points)  for which the absolute magnitudes are converted to bolometric luminosity by applying the bolometric correction of \cite{Lyman2014}. For comparison, we also show the bolometric light curve of SN1987A from \cite{Suntzeff1990} (red solid curve),  $^{56}\text{Co} \rightarrow ^{56}\text{Fe}$ decay rate (dashed blue curve), and $^{56}\text{Ni} \rightarrow ^{56}\text{Co}$ decay rate (dash-dotted green curve). Note that the decay rates shown are relative and only represent the conversion slope of the radioactive chain. The magenta line shows a linear fit to the tail bolometric luminosity. We estimate $M_{\text{Ni}} = 0.10\pm0.01 M_\Sun$ from the tail luminosity of \kspn. }
\label{fig:lbol}
\end{figure}

\subsection{Bolometric Light Curve and $^{56}\text{Ni}$ mass}
\label{sub:lbol}
Figure~\ref{fig:lbol} shows the evolution of the bolometric luminosity (filled black circles) of \kspn\ in comparison with
that of SN1987A (red solid curve) as well as the relative decay rates of  $^{56}\text{Co} \rightarrow ^{56}\text{Fe}$ (dashed blue curve) and $^{56}\text{Ni} \rightarrow ^{56}\text{Co}$ (dash-dotted green curve). In order to obtain bolometric luminosities of \kspn, the absolute $V$-band magnitudes of \kspn, shown in Figure~\ref{fig:comp}, are converted to bolometric magnitudes using
${\rm BC}=M_{\rm bol}-M_V$, where BC is the bolometric correction (BC) of \cite{Lyman2014}. In this formalism, BC is a color-dependent (here, we use $V-I$ index) polynomial with coefficients listed in their Table 3 and 4. Finally, bolometric magnitudes are converted to luminosities assuming M$_{\text{bol},\Sun}=4.74$ mag and L$_{\text{bol},\Sun}=3.83 \times 10^{33}$ erg s$^{-1}$. Since the bolometric luminosity depends on $V-I$ color evolution, our bolometric light curve is cut off at $\sim$130 days when \kspn\ goes below the detection limit in $V$ band.

The $^{56}$Ni mass ($M_{\text{Ni}}$) synthesized in Type II SNe is often estimated by the tail bolometric luminosity, as the tail is dominantly powered by the $ ^{56}\text{Ni} \rightarrow ^{56}\text{Co} \rightarrow ^{56}\text{Fe}$ radioactive decay chain. Figure~\ref{fig:lbol} indicates that for $t\gtrsim$ 115 days the bolometric light curve of \kspn\ transitions to a linearly declining radioactive tail which is $\sim25\%$ more luminous than that of SN1987A, implying a higher synthesized $M_{\text{Ni}}$ based on the direct comparison of luminosities \citep[See Equation~3 of][]{Valenti2016}. Following Appendix A of \cite{Valenti2008}, we estimate $M_{\text{Ni}} = 0.10\pm0.01 M_\Sun$, as presented in Table~\ref{tab:gen_params}, using the known decay times of $^{56}$Ni and $^{56}$Co as well as their associated energy generation rates assuming the complete trapping of $\gamma$-rays and positrons (supported by the  similar slopes of the radioactive tail and $^{56}\text{Co} \rightarrow^{56}\text{Fe}$ decay rate, as shown in Figure~\ref{fig:lbol}).




\begin{figure*}[!t]
\centering
\includegraphics[trim=1cm 1cm 1cm 0.9cm, clip=true,scale=0.76]{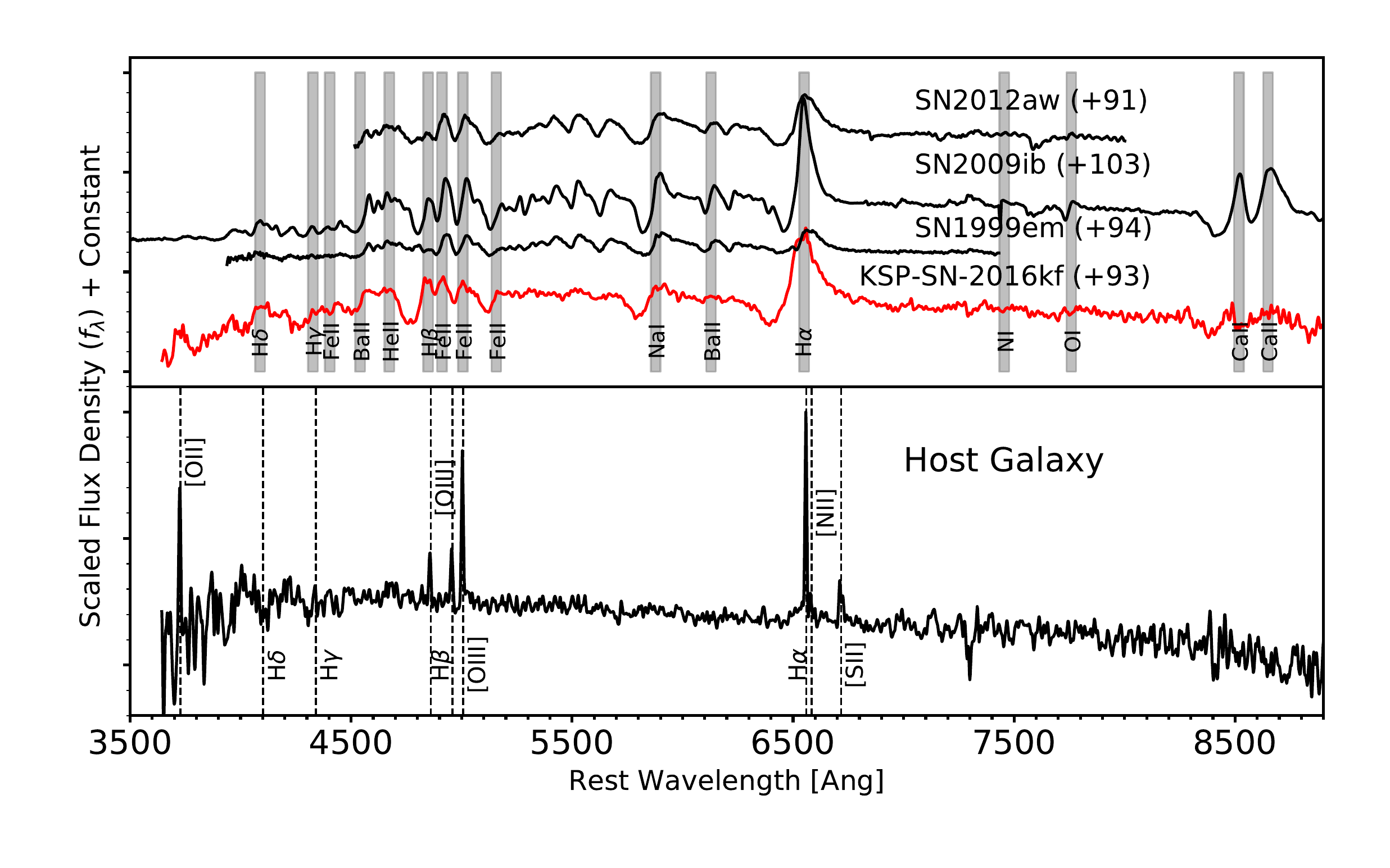}
\caption{\emph{Top:} Comparison of \kspn\ (red curve) and other H-rich Type II SNe (black curves) at similar epochs. Note that the flux density for each SN is scaled by its distance squared for the fair comparison of features. The gray vertical lines indicate the wavelength of typical spectral lines visible at this epoch. \emph{Bottom:} the spectrum of \kspn's host galaxy. The emission lines are marked with dashed lines. 
}
\label{fig:spec}
\end{figure*}

\section{Spectroscopy of \kspn}
\label{sec:spec}

Figure~\ref{fig:spec} (top panel) compares the spectrum of \kspn\ taken at an epoch of 93 days, 
which is near the transition from the plateau phase to the radioactive tail phase, with those of the other three well-observed H-rich Type II SNe at a similar phase. The spectrum of \kspn\ exhibits prominent P-Cygni Balmer lines, classifying \kspn\ as a Type II SN. Using the redshift of $z=0.043 \pm 0.002$ obtained from the host galaxy (see $\S$\ref{sub:host}), the spectrum is de-redshifted to the rest frame.  


In order to obtain the expansion velocities of \kspn, we fit a Gaussian profile to the absorption feature of the P-Cygni profile of four Balmer lines of H$\alpha$, H$\beta$, H$\gamma$ and H$\delta$ as well as prominent iron lines, \ion{Fe}{2}$\lambda \lambda 4924$, \ion{Fe}{2}$\lambda \lambda 5018$, and \ion{Fe}{2}$\lambda \lambda 5169$ and measure the expansion velocity of each line at the minimum absorption of the fitted Gaussian profile. The measured expansion velocities, listed in Table \ref{tab:lines}, have a wide range of values; the H$\alpha$ has the highest velocity at $\sim$7960 km~s$^{-1}$, while \ion{Fe}{2}$\lambda \lambda 4924$ has the lowest velocity among the selected lines at $\sim$2798 km~s$^{-1}$. We can see that all Balmer lines have a higher expansion velocity than any of the three \ion{Fe}{2} lines. This is because H lines are formed at much lower optical depths than \ion{Fe}{2} lines, hence exhibit higher velocities. The velocities of \ion{Fe}{2}$ \lambda \lambda5018$ and \ion{Fe}{2}$\lambda \lambda5169$ are often associated with the photospheric expansion velocity, as they are formed closer to the photosphere \citep{Takats2012}.


\begin{deluxetable}{lcc}
\tabletypesize{\footnotesize}
\tablecolumns{2} 
\tablewidth{0.99\textwidth }
 \tablecaption{Line velocities at epoch of 93 days\label{tab:lines} }
 \tablehead{
 \colhead{Line \& Wavelength}\hspace{0.8cm} & \colhead{Velocity [km~s$^{-1}$]} 
 } 
\startdata 
\hspace{0.3cm}H$\alpha \lambda\lambda 6563$ \hspace{0.8cm}& 7960 $\pm$ 60 & 
 \\
\hspace{0.3cm}\ion{Fe}{2}$\lambda \lambda5169$ \hspace{0.8cm}& 3192 $\pm$ 63 &
\\
\hspace{0.3cm}\ion{Fe}{2}$ \lambda \lambda5018$ \hspace{0.8cm}& 3134 $\pm$ 84 &
\\
\hspace{0.3cm}\ion{Fe}{2}$ \lambda \lambda 4924$ \hspace{0.8cm}& 2798 $\pm$ 55 &
\\
\hspace{0.3cm}H$\beta \lambda \lambda4861$ \hspace{0.8cm} & 5860 $\pm$ 45&
 \\
\hspace{0.3cm}H$\gamma \lambda \lambda4340$ \hspace{0.8cm}&  5452 $\pm$ 167  &
\\
\hspace{0.3cm}H$\delta  \lambda \lambda4101$\hspace{0.8cm}& 4818 $\pm$ 141 & 
\enddata
\tablecomments{The uncertainties are associated with the profile and continuum fitting method used for finding the minimum of the absorption features.}
\end{deluxetable}

\begin{figure}[!t]
\centering
\includegraphics[trim=0.5cm 1cm 0.5cm 1cm, clip=true,scale=0.55]{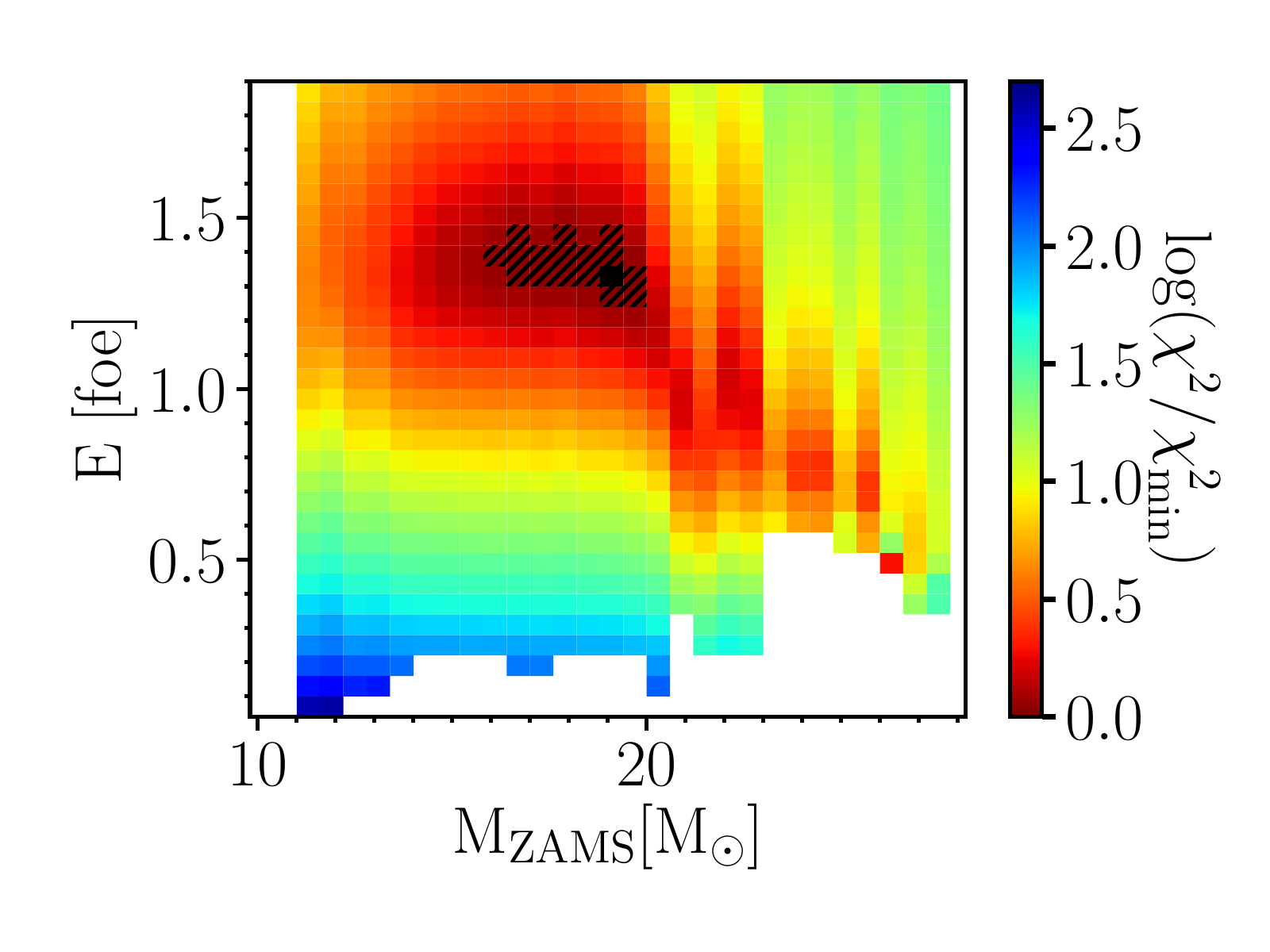}
\caption{ The scaled goodness of fit, $\log (\chi^2/\chi_{\min}^2)$, where $\chi_{\min}=5100$ represents the goodness of fit for the best-fit model, for a grid of SNEC simulations with 5~M$_\Sun$ $^{56}$Ni mixing scheme and different ZAMS masses and explosion energies. The hatched squares represent best 2-percentile of fitted models. The best-fit model is marked with a filled black square and has $\sim$18.8 M$_\Sun$ ZAMS mass, $\sim$1.3 foe explosion energy, and 5~M$_\Sun$ $^{56}$Ni. The evolved RSG progenitor of this model has M~$\simeq15 ~ \rm{M}_\Sun$ and R~$\simeq1040~ \rm{R}_\Sun$. }
\label{fig:snec}
\end{figure}

Although the spectrum of \kspn\ in Figure~\ref{fig:spec} (top panel) is broadly similar to those of the other H-rich Type II SNe
that we compare with, there are a few notable differences: (1) the H$\alpha$ line resembles that of SN2012aw, but broader (in both emission and absorption) with a shallower absorption feature, (2) the H$\beta$ line appears much stronger than that of other SNe, producing a deep and broad absorption trough redwards of a prominent emission peak, which could be due to blending with other nearby lines, and (3) the \ion{Ba}{2} line has a very small emission peak with no notable trough and is the weakest compared to that of other H-rich Type II SNe.

For constraining the metallicity of the SN progenitor, we measure the pseudo-Equivalent Width (pEW) of the absorption feature of \ion{Fe}{2}$ \lambda \lambda5018$ line. Recently, it has   been shown that the pEW of this line during the plateau phase  can probe the progenitor metallicity of H-rich Type II SNe. \citep{Dessart2014,Taddia2016b, Gutirrez2018}. According to Figure 1 of \cite{Taddia2016b}, pEW of \ion{Fe}{2}$ \lambda \lambda5018$ is correlated with the SN phase and metallicity. Our measured pEW of $\sim$18 \AA\ at the epoch of 93 days corresponds to the metallicity of 0.1--0.4~$Z_\Sun$. This metallicity measurement complements the metallicity estimated based on several line diagnostics of the host galaxy spectrum (see $\S$\ref{sub:host})


\section{Host Galaxy of \kspn}
\label{sub:host}

Figure~\ref{fig:spec} (bottom panel) shows the spectrum of the galaxy identified with ``G'' in Figure \ref{fig:host}. The spectrum exhibits prominent Balmer H emission lines including H$\alpha$ and H$\beta$ as well as several other lines such as [\ion{O}{3}]~$\lambda \lambda$3727, [\ion{O}{3}]~$\lambda \lambda$4959,[\ion{O}{3}]~$\lambda \lambda$5007, [\ion{N}{2}]~$\lambda \lambda$6584, and [\ion{S}{2}]~$\lambda \lambda$6717. 
The redshift of the galaxy, measured from the H$\alpha$ line, is $z \simeq 0.043$,
which is equivalent to the luminosity distance of 197.4~Mpc (see \S~\ref{sec:phot}.3). 
This redshift is consistent with that of \kspn\ as shown in \S~\ref{sec:spec}, 
confirming that the galaxy is the host galaxy of the SN.

The host galaxy is 4\arcsec\ (or $\sim$3.4 kpc at the angular diameter distance of 181.5~Mpc) away from the SN in the eastern direction. 
It has a small effective radius of $\sim$ 2.2~kpc (or $\sim$2\farcs5). The galaxy is relatively bright with absolute magnitude of $-17.57 \pm 0.06$ ($B$), $-18.17 \pm 0.08$ ($V$), and $-19.05 \pm 0.06$ ($I$) mag for the size of the galaxy. 
Using the FAST stellar population synthesis code \citep{Kriek2009} we find a best-fit stellar mass for the host galaxy of  $\log$ (M/M$_\odot$) $=$ 8.73$^{+0.51}_{-0.73}$. 
These model calculations are based on the stellar library of \citet{Maraston2005} 
and the assumption of exponential star formation history and Salpeter IMF.

The observed line intensity ratio between H$\alpha$ and H$\beta$ lines integrated
over the host galaxy is $3.11 \pm 0.26$ (Figure~\ref{fig:spec}). 
Within the temperature range of 5000--20000~K, the intrinsic line intensity ratio 
of the two lines changes between 3.04 and 2.75 \citep{Dopita2003} 
for electron number density of 100 cm$^{-3}$. 
By comparing the observed and intrinsic ratios, we obtain the extinction
$E(B-V)$ in the range of 0.02--0.10~mag \cite[see][and references therein]{Groves2012},
or $A_V$ = 0.07--0.33 mag for $R_V=3.1$ using the extinction correction model by \citet{Fitzpatrick1999}. 
Considering that \kspn\ is located substantially away from the host galaxy (Figure~\ref{fig:host}), 
this small estimated extinction toward the center of the host galaxy 
is consistent with a negligible host-galaxy extinction toward the SN (see $\S$\ref{sub:red}).

We find the metallicity of the host galaxy of \kspn\ to be sub-solar, i.e., $Z/Z_\Sun \simeq 0.4$, based on
the analysis of the observed line ratios sensitive to O abundance and metallicity. 
For this estimate, we use O3N2 and N2 line ratio indicators \citep{Marino2013} 
obtained with the observed fluxes of H, O and N lines in the host galaxy spectrum (Figure~\ref{fig:spec}).
The O3N2 and N2 indicators provide the O abundance of $8.23 \pm 0.03$ and $8.24 \pm 0.06$, respectively, for the host galaxy,
corresponding to to $Z/Z_\Sun = 0.4 \pm 0.1$ assuming $\rm 12+\log({\rm O/H})_\Sun=8.69$ \citep{Asplund2009}. 
All abundances are calculated in \texttt{PyMCZ} code with the observed line fluxes \citep{Bianco2016}. 
The abundances from O3N2 and N2 line indicators are
$\sim$0.22 and $\sim$0.21 dex, respectively, 
smaller than the median value of the O abundance found in other Type II SNe \citep{Anderson2016}. 
The $Z/Z_\Sun \simeq 0.4$ host galaxy metallicity coincides with the upper bound value of
the SN metallicity, i.e., $Z/Z_\Sun \simeq 0.1-0.4$ (\S\ref{sec:spec}).
We note that this is consistent with the location of \kspn\ in the outskirts of the host galaxy
since the galaxy metallicity tends to drop from the center to outskirts \citep{Taddia2015}.

\begin{figure}[!t]
\centering
\includegraphics[trim=0.75cm 2.5cm 0.3cm 1.3cm, clip=true,scale=0.65]{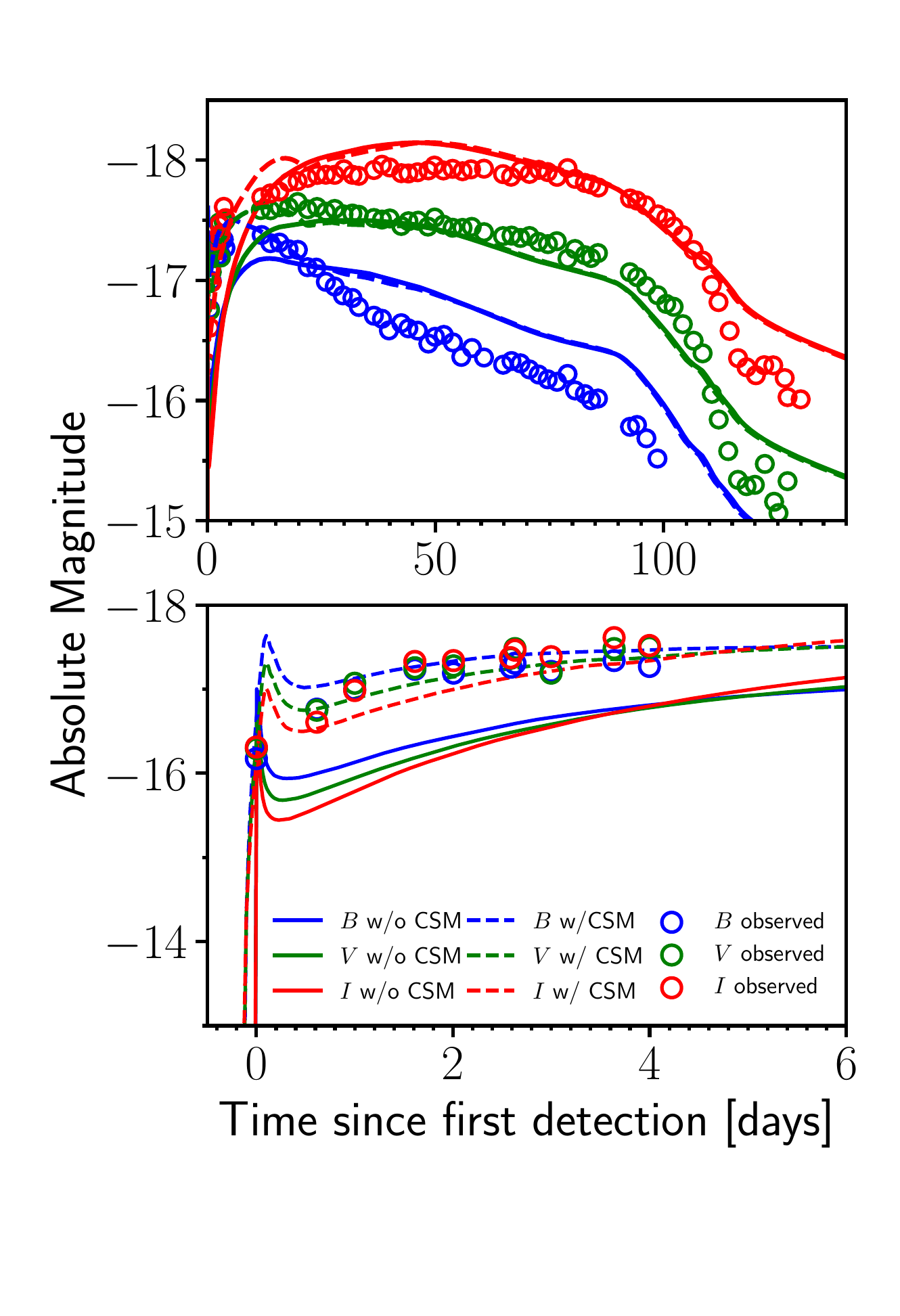}
\caption{Comparison of the light curves for 1) the best-fit model without a CSM component (solid curves), 2) the best-fit model with a dense CSM component (dashed curves), and 3) observed values (open circles). \emph{Top:} the observed and simulated light curves of \kspn. \emph{Bottom:} same as top panel but zoomed-in to show the early rise epochs. }
\label{fig:lccsm}
\end{figure}


\section{Progenitor and Explosion Parameters}
\label{sec:prog}

\subsection{Hydrodynamic Modelling}
\label{sub:hydro}
To constrain the physical parameters of \kspn, we conduct  the radiation transfer hydrodynamical simulations that give the best-fit parameters for the light curves. The simulations are carried out in open-source code \texttt{SNEC} \citep{Morozova2015}, a 1D flux-limited radiation transfer code which treats ionization levels and recombination under the local thermodynamic equilibrium (LTE) assumption. Similar to \cite{Morozova2018}, we construct a grid of simulations with non-rotating RSG progenitor models in the ZAMS mass range of 11--28~M$_\Sun$ with 0.6~M$_\Sun$ resolution created by the \texttt{KEPLER} stellar evolution code \citep{Woosley2007}. The explosion is launched by the thermal bomb mechanism and is evolved for 120 days, i.e., before the onset of the optically thin phase. We excise 1--2 M$_\Sun$ at the center of the explosion to account for the mass of the proto-neutron SN remnant  \citep{Morozova2018}.  We adopt \cite{Paczynski1983} equation of state implemented in \texttt{SNEC} for the RSG model and set the numerical grid size to 1000 cells in all of our simulations. In addition to varying the ZAMS mass of the \texttt{KEPLER} progenitor models, we also vary the explosion energy in the range of 0.25--1.96 foe (where 1 foe $=10^{51}$ erg~s$^{-1}$) with 0.06 foe resolution. For  $^{56}$Ni mixing, we adopt three schemes for which the $^{56}$Ni is roughly mixed in the progenitor up until 3~M$_\Sun$, 5~M$_\Sun$, and 7~M$_\Sun$ in mass coordinate within the progenitor. As described in $\S$\ref{sub:lbol}, we fixed M$_{\text{Ni}}= 0.10 M_\Sun$ in the simulations.

The $BVI$ light curves from \texttt{SNEC} simulations are then compared to the observed light curves by calculating $\chi^2$ as below,
\begin{eqnarray}
\label{eq:goodness}
\chi^2 &=& \mathlarger{\mathlarger{\mathlarger{\sum}}}_{\lambda \in \{V,I\}} \ \ \mathlarger{\mathlarger{\mathlarger{\sum}}}_{t_{\text{SBO}} \leq t_*\leq 105} \frac{ \Big[M_{\text{obs},\lambda}(t_*) - M_\lambda(t_*)\Big]^2} {\Big[\sigma_{\text{obs},\lambda}(t_*)\Big]^2},
\end{eqnarray}
where $\lambda$ denotes the corresponding band, $t_*$ the time of observation, $M_{\text{obs},\lambda}$ the observed absolute magnitudes, $M_{\lambda}$ the simulated absolute magnitudes, and $\sigma_{\text{obs},\lambda}$ the uncertainty of the observed magnitudes. We exclude the $B$-band light curve data due to the non-negligible effect of iron group line blanketing to the observed brightness \citep{Dessart2010}. We also exclude data from $t_*>105$ days because the LTE assumption may not hold for those epochs. 

Figure~\ref{fig:snec} shows the normalized $\chi^2$ by minimum $\chi_{\min}^2$ value from our \texttt{SNEC} simulations  as a function of explosion energy (E) and M$_\text{ZAMS}$ for the case of 5~M$_\Sun$ $^{56}$Ni mixing. We found that 5~M$_\Sun$ $^{56}$Ni mixing scheme fits the observed light curves much better than the other two mixing schemes. The hatched squares represent best 2 percentile of best-fit models. The E and M$_\text{ZAMS}$ of these models are, respectively, in the range of 1.24--1.4 foe and 15.8--19.4~M$_{\Sun}$, among which a model with 1.3~foe  and 18.8~M$_\Sun$ ZAMS mass best fits the observed light curves and is marked with a filled black square. The evolved RSG progenitor of this model has mass $M\simeq15~\rm{M}_\Sun$ and radius $R\simeq1040~\rm{R}_\Sun$.  Figure 9 compares the light curves of the best-fit model with the observed light curves of \kspn. The model closely fits the observed values during the late time evolution (i.e., $t>20$ days) in $V$ and $I$ bands, but during the early epochs the fits underestimate the luminosity of the observed light curve, probably due to the presence of CSM (see below).

\begin{figure}[!t]
\centering
\includegraphics[trim=0.5cm 1.5cm 0.5cm 1.3cm, clip=true,scale=0.62]{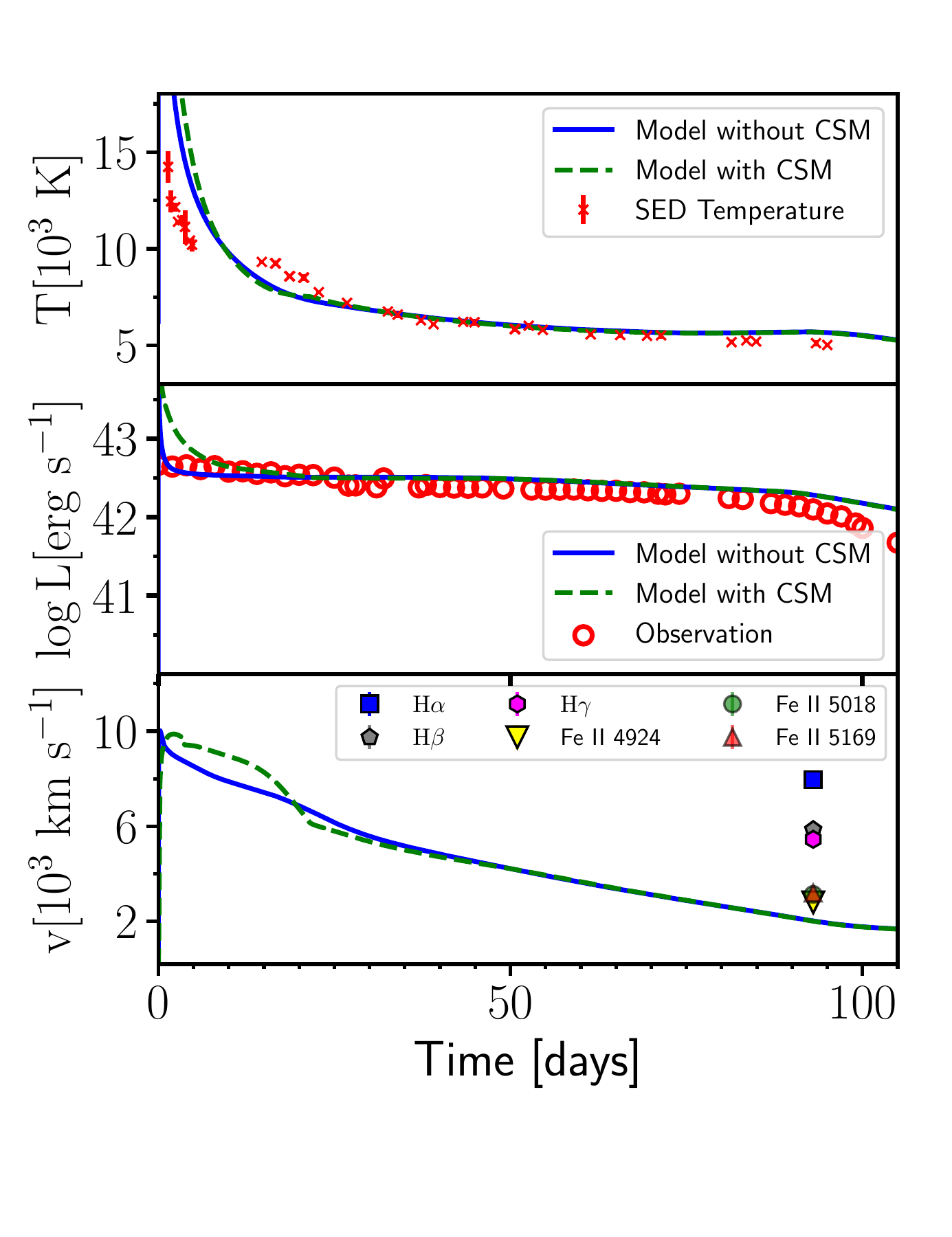}
\caption{Blackbody temperature, bolometric luminosity, and photospheric velocity for the best-fit model without CSM component (solid blue curve) and with dense CSM component (dashed green curve) as well as the corresponding observationally estimated values. \emph{Top:} photospheric temperature over time. Red open circles represent the blackbody temperature for the observed $BVI$ magnitudes. \emph{Middle:} bolometric light curve. The observed luminosities are shown with red open circles. \emph{Bottom:} photospheric velocity. Several line velocities are also plotted. }
\label{fig:csm}
\end{figure}

\subsection{Potential Interaction with Dense CSM} 
\label{sec:csminter}
As shown in Figure 9, the comparison between  the best-fit light curve and the observed light curves indicate the existence of additional emission in the early epochs. In order to understand the nature of this emission, we conducted additional  \texttt{SNEC} simulations including a CSM component. We adopt a dense optically-thick, constant-velocity CSM with the density profile $\rho(r)={\rm K}/r^2$ for $r<\rm{R}_\text{CSM}$ \citep[e.g.,][]{Morozova2018} around the best-fit progenitor model described above. In the simulations, we vary the parameter K in the range of $11 \times 10^{17}-3 \times 10^{18}$~g~cm$^{-1}$ with resolution of $10^{17}$~g~cm$^{-1}$ and $\rm{R}_\text{CSM}$ in the range $1040-3800$~R$_\Sun$ with resolution 100~R$_\Sun$. We assume the composition and temperature of the CSM are equal to the outer part of a RSG with H fraction of 0.61, He fraction of 0.37, and temperature of 2400~K. The adoption of this type of dense optically-thick CSM is motivated by the increasing  evidence of RSGs undergoing strong mass-loss shortly prior to the collapse of the core \citep[see e.g.,][]{Khazov2016, Yaron2017, Hosseinzadeh2018, Bullivant2018, Forster2018}. 


We select the best-fit model with CSM component using Equation \ref{eq:goodness} in a similar manner as in $\S$\ref{sub:hydro} and find that a model with R$_\text{CSM}\simeq1640~ \rm R_\Sun$ and K$\simeq4 \times 10^{17}$~g~cm$^{-1}$ best matched observations. With these values of K and R$_\text{CSM}$, the total CSM mass between the progenitor and R$_\text{CSM}$ is $0.11$~M$_\Sun$, and this CSM mass along with the above K is equivalent to the mass loss rate of $\sim$0.08~M$_\Sun$~yr$^{-1}$ assuming typical wind velocity of 10~km~s$^{-1}$. The SNEC simulations also provide information for the SBO, which occurs inside the CSM near its outer edge at radius $\sim$1550~R$_\sun$ or optical depth $\tau \simeq 95$. Figure~\ref{fig:lccsm} also shows the light curves of the best-fit model with the CSM component along with the best-fit model without the CSM component. As in the figure, the model that includes the CSM component fits the observed early light curves significantly better than the best-fit model without it.  

Figure~\ref{fig:csm} compares observationally estimated blackbody temperatures (top panel) and bolometric luminosities (middle panel) with those of the best-fit models, both with and without the CSM component. The temperatures represent best-fit black body temperatures from the observed $BVI$ magnitudes, while the observed bolometric luminosity is calculated  after applying proper bolometric corrections ($\S$\ref{sub:lbol}). The modelled bolometric luminosity is measured at the photosphere. Figure~\ref{fig:csm} (bottom panel) also compares velocities of several spectral lines at the epoch of 93 days (presented in Table~\ref{tab:lines}) and the simulated photospheric velocities. 


As seen in Figure \ref{fig:csm}, the temperature and luminosity derived from observations
match well with those from the best-fit model during
the later epoch. There are, however, slight discrepancies in temperature
and luminosity during the early ($t\lesssim5$ days) epoch:
(1) in temperature, the models (both with and without CSM) predict
higher (by 2500~K) temperature than the observations;
(2) in luminosity; while the observations match the predictions from the model without the CSM component,
the model with CSM overpredicts the observed luminosity.
The discrepancy (1) can be due to inaccurate temperatures estimated from $BVI$ light curves as the temperature is high and the observed bands do not cover the peak SED intensity or potential deviation from a blackbody spectrum in the presence of CSM \citep{Chevalier2011}, while the apparent discrepancy in (2) may arise from inaccurate bolometric correction for the early envelope cooling phase for SNe with CSM.  These overall agreement in temperature and luminosity, together with the similarity in the photospheric velocities of the model and \ion{Fe}{2} line velocities---which is associated with the photospheric velocity as discussed by $\S$\ref{sec:spec}--- indicate that the observed light curves of \kspn\ are reasonably well-reproduced in our simulations. 

It is worth noting that our estimates on the K and R$_\text{CSM}$ of the potential dense CSM component should be regarded as a lower limit due to the LTE assumption in \texttt{SNEC}. It is likely that this assumption breaks down when the shock moves into the CSM, in which case the temperature of the photons would significantly exceed that calculated in the simulations under the LTE assumption. This  leads to a longer rise time as it takes more time for the color temperature (i.e., spectrum peak) to drop into the observed band \citep{Nakar2010}.

\begin{figure}[!t]
\centering
\includegraphics[trim=0cm 0cm 0cm 0cm, clip=true,scale=0.47]{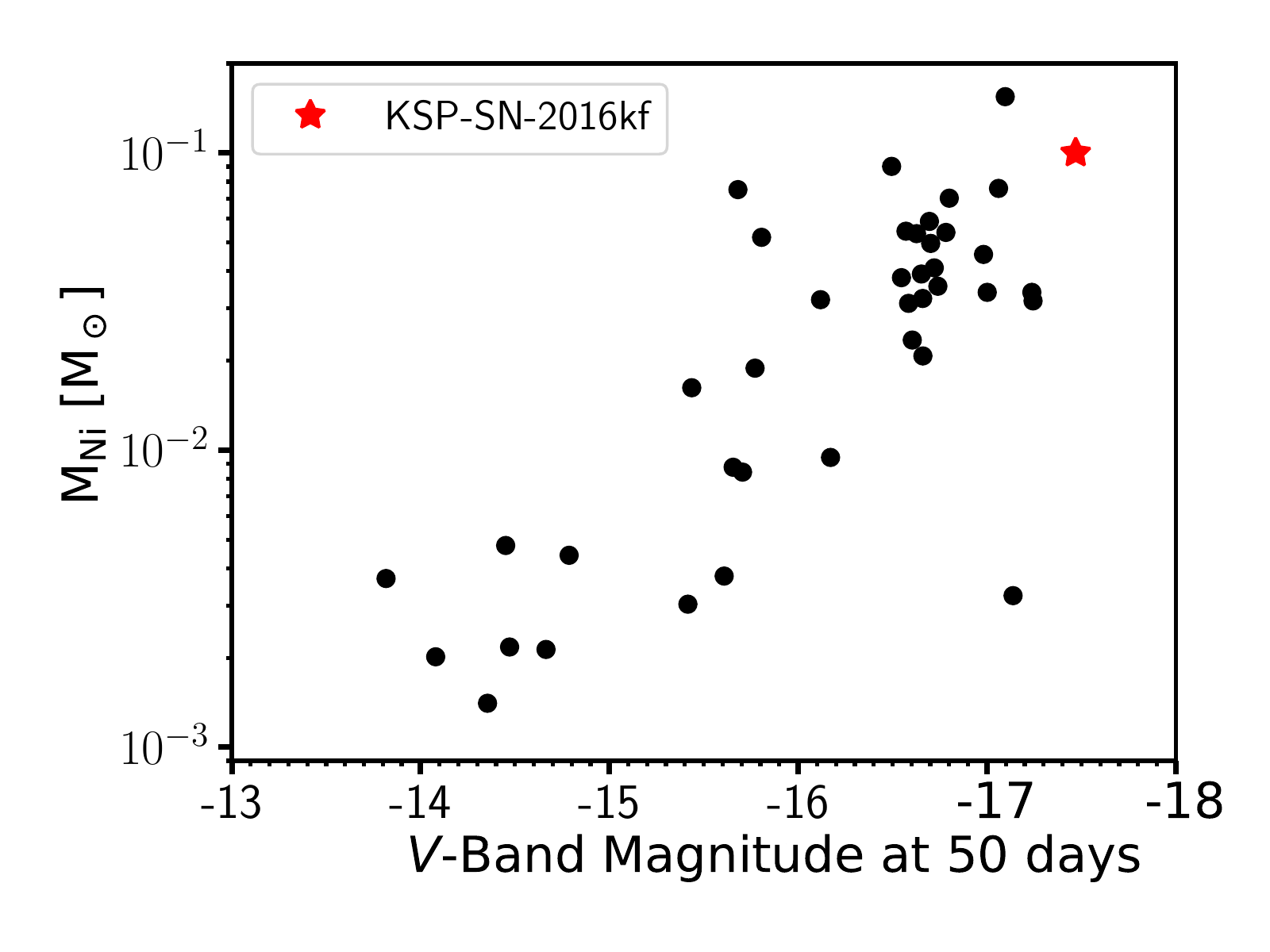}
\caption{$^{56}$Ni mass and plateau magnitude at the epoch of 50 days for a sample of Type II SNe in \cite{Valenti2016}. The red star represents \kspn\ which is located at higher end of both $^{56}$Ni mass and plateau magnitude. 
}
\label{fig:ni56}
\end{figure}


\begin{figure}[t]
\centering
\includegraphics[trim=0cm 0cm 0cm 0cm, clip=true,scale=0.47]{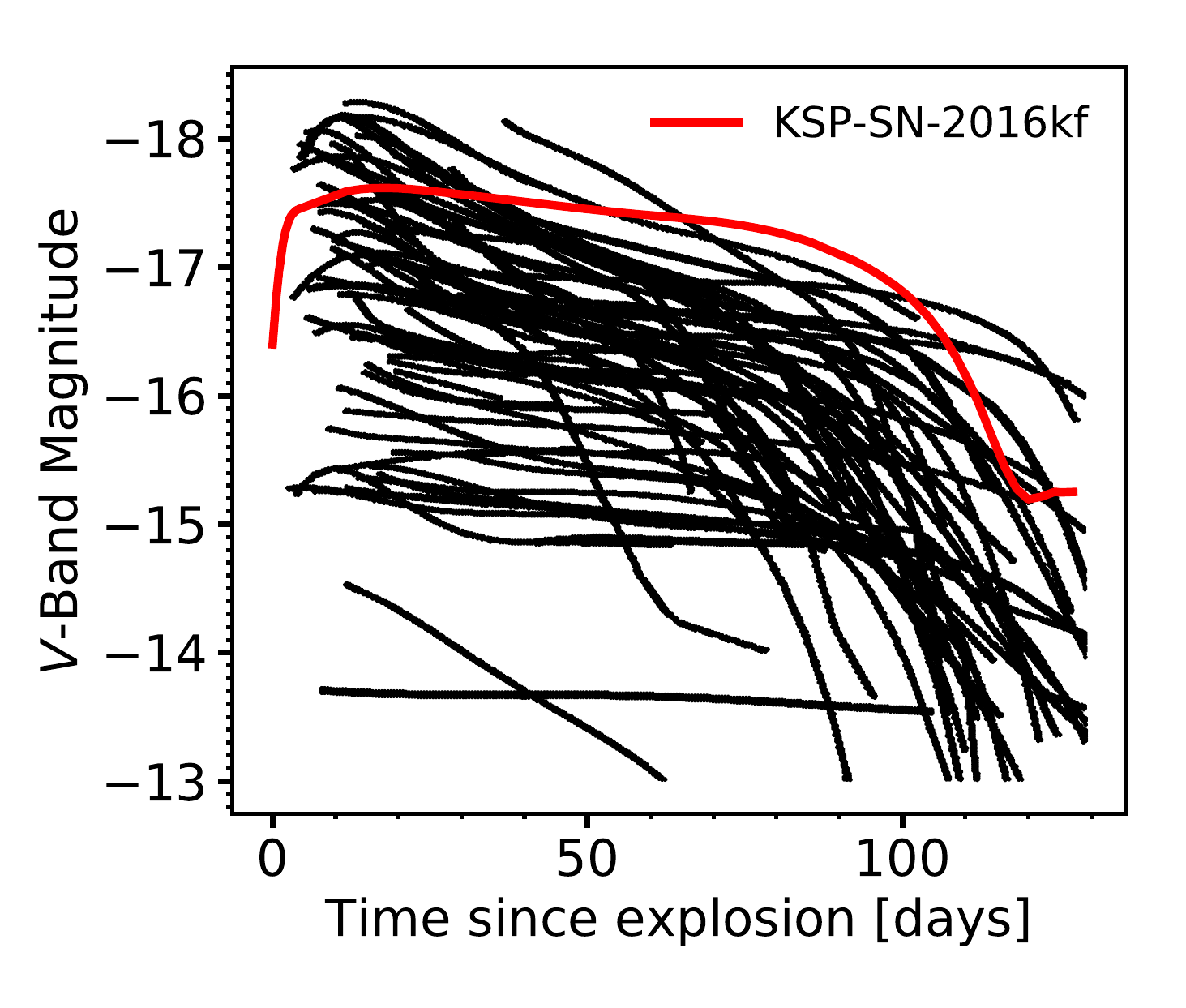}
\caption{Comparison of the $V$-band light curve of \kspn\ (red curve) and  that of other Type II SNe in \cite{Anderson2014}. The light curve of \kspn\ is comparatively flat and luminous during the plateau phase.}
\label{fig:compare}
\end{figure}



\section{Discussion}
\label{sec:diss}
\kspn\ is a H-rich Type II SN with a long rise time ($t_\text{rise} \simeq 19.9$ days in $V$ band), a relatively luminous peak ($M_V \simeq-$17.62~mag), a highly luminous plateau ($M_V \simeq -$17.47~mag at the epoch of 50 days), and a low decay rate  ($s \simeq 0.53$~mag per 100 days). Its tail part ($t\gtrsim 105$~days) is exceptionally bright, requiring a large $^{56}$Ni mass of $ \sim 0.1$~M$_\sun$ to power the tail luminosity. The SN likely resulted from an  explosion with  $\sim$1.3~foe energy in a RSG, evolved from a $\sim$ 18.8~M$_\sun$ ZAMS mass, whose mass and radius are $\sim 15$~M$_\sun$ and $\sim 1040$~R$_\sun$, respectively, at the time of explosion. The early light curves of \kspn\ indicate the presence of CSM. 

Figure~\ref{fig:ni56} compares the $^{56}$Ni mass and the plateau magnitude of \kspn\ at the epoch of 50 days to those of most other Type II SNe at the same epoch presented in \cite{Valenti2016}, showing a positive correlation between two parameters. This correlation has been also reported in several previous studies \citep{Hamuy2003,Spiro2014,Pejcha2015,Valenti2016}. \kspn\ appears to have the most luminous plateau among the shown sample of Type II SNe as well as one of the highest inferred $^{56}$Ni masses, while it still follows the known correlation. The $^{56}$Ni mass of Type II SNe has also been known to be correlated with 
the explosion energy \citep{Hamuy2003,Pejcha2015a,Muller2017}, and the $^{56}$Ni mass of $\sim 0.1$~M$_\sun$ and 1.3~foe explosion energy of \kspn\ appear to follow the correlations in those studies.


Figure~\ref{fig:compare} shows the $V$-band light curves of a sample of Type II SNe presented in \cite{Anderson2014} in comparison with that of \kspn. We can easily identify that the plateau of \kspn\ is flatter than those of other SNe with comparable luminosity. The known correlation between the high peak luminosity and the plateau decay rate \citep{Anderson2014} predicts much higher decay rate than what we observed for \kspn. Since we estimated a large amount of $^{56}$Ni, its flatter plateau may indicate the presence of an early contribution of the $^{56}$Ni radioactive decay to the plateau luminosity \citep{Nakar2016}.




As in Figure~\ref{fig:ni56}, \kspn\ has one of the largest $M_{\rm{Ni}}$ among the sample of \cite{Valenti2016}. This is also consistent with the sample of 19 H-rich Type IIP SNe in \cite{Muller2017}, whose median $M_{\rm{Ni}}$ is 0.031 M$_\Sun$. High $^{56}$Ni masses in the range of 0.1--1.6 M$_\Sun$ have been previously reported for several Type II SNe including SN1992H \citep{Hamuy2003}, SN1992af \citep{Valenti2016}, SN1992am \citep{Hamuy2003}, SN 2016ija \citep{Tartaglia2018}, and ASASSN-15nx \citep{Bose2018}, although there are large uncertainties associated with the inferred large $M_{\rm{Ni}}$ of some of these SNe due to possible alternative power sources at later epochs, unknown dust extinction, and limited photometric data over the light curve tail.  

It appears that the observationally estimated $^{56}$Ni mass of a group of SNe, including those mentioned above, is higher than the typical $M_{\rm{Ni}}$ upper limit of $\sim$0.12~\Msolar\ predicted by neutrino-driven CCSN simulations \citep[e.g.,][]{Ugliano2012,Pejcha2015a, Sukhbold2016}. Therefore, it is important to increase the sample size of SNe with large $^{56}$Ni mass to draw firm conclusions on the compatibility of neutrino-driven CCSN simulations and the observations in terms of $^{56}$Ni mass production. For \kspn, our estimated $M_{\rm{Ni}} \simeq 0.1$~M$_\sun$ is within the predicted $M_{\rm{Ni}}$ range of 0.003--0.12~M$_\Sun$ for the successful SN explosions in the neutrino-driven CCSN simulations of \cite{Sukhbold2016} and our obtained explosion energy of 1.3~foe is compatible with the production of the 0.1 M$_\Sun$ of $^{56}$Ni based on the positive correlation between two parameters \citep[][see their Figure 17]{Sukhbold2016}. 

The large ZAMS mass, $\sim 18.8$~\Msolar\ (Section \ref{sec:prog}), of the progenitor of \kspn\
provides an important insight into how H-rich Type II SNe explode. According to \cite{Smartt2009}, large RSGs, whose mass during their ZAMS phase were greater than $\sim$17 \Msolar, have been very rare as progenitors of Type II SNe.
This lack of RSG progenitors with a large ZAMS mass, known as the ``RSG problem'', appears to be 
consistent with what is predicted by numerical simulations of neutrino-driven CCSNe which substantially favor ZAMS stars of smaller mass, $\lesssim 17$~\Msolar, as the progenitor 
of Type II SNe \citep{Ugliano2012,Oconnor2013,Sukhbold2016}. In this scenario, ZAMS stars of mass $\gtrsim 17$~\Msolar tend to directly implode to a BH or produce failed SNe without SN-like activity. The recent discovery of a large mass RSG progenitor with metallicity of $\sim 0.1$ \citep{Anderson2016} shows that metallicity may play an important role, although it is difficult to reach a firm conclusion due to the lack of other observational examples supporting the idea.
\kspn\ appears to support the interpretation that a link between low metallicity progenitors and SN in massive RSGs may exist.

The light curves of \kspn\ indicate the presence of a CSM component similar to the one assumed in \cite{Morozova2018}. However, recent numerical simulations including wind accelation predict a CSM component that is more extended and less massive than what is required in \kspn.
This emphasizes the need to obtain early spectra that can constrain the properties of CSM more rigorously.

Finally, we note that our results of the progenitor parameters of \kspn\
are based on model calculations relying on three key assumptions:
(1) the SN explosion is spherical; hence, a 1D model is sufficient; 
(2) the LTE assumption holds for the first 105 days since the explosion; and 
(3) non-rotating pre-SN models are adequate for the progenitor.
Although these assumptions have been commonly adopted in similar studies of Type II SNe in practice, 
the results based on these assumptions should be taken with caution in principle
since we cannot rule out the possibility that the real situations
are in fact much different from the assumptions. 
The first assumption on the sphericity of the SN progenitor and explosion 
is motivated by the fact that developing significant asymmetries and non-radial flows in RSGs 
is much more difficult than in compact stripped-envelope progenitors and requires extreme aspherical explosions or oblate progenitors \citep[e.g.]{Matzner2013}. 
Similarly, the LTE assumption is relevant for RSG progenitors as the post-SN RSG shock propagates slowly enough for the shocked material to generate sufficient photons for maintaining the equilibrium \citep{Nakar2010}, although departure from LTE may happen considering CSM interaction as discussed in \S\ref{sec:csminter}. For the third assumption on non-rotating RSG models, 
it is important to note that fast rotation may cause the H envelope of the stars in the initial mass range of 15--20 \Msolar to be removed due to changes in core structure and enhanced mass loss, leading to Type IIb or Ib/c SNe rather than Type II SNe \citep{Hirschi2004}. It is, therefore, more difficult for rapidly-rotating and massive RSGs to explode as a Type II SN.

\section{Summary and Conclusion}
\label{sec:sum}
In this paper, we present the photometric, and spectroscopic analyses of \kspn\ and its host galaxy discovered by KMTNet. Our multi-color high-cadence observations of \kspn\  give a tight constraint on the shock breakout epoch as well as reliable estimates of its rise times. Based on a comprehensive modelling of the observed properties and also comparison with other Type II SNe, we summarize the properties and  peculiarities of \kspn\ as follows: \\
(1) \kspn\ has an exceptionally long rise time for a H-rich Type II SN, particularly among high luminosity ones, with $t_\text{rise}\simeq20$ and $\simeq 50$~days in $V$ and $I$ bands, respectively.\\
(2) The light curves of \kspn\ have a moderately bright peak with $M_V\simeq -17.6$~mag, but highly luminous plateau with $M_V\simeq -17.4$~mag at 50 days. The plateau is one of the most luminous ones ever observed for H-rich Type II SNe.\\
(3) \kspn\ also has an exceptionally luminous radioactive tail, requiring $0.10\pm0.01$~M$_\sun$ of $^{56}$Ni mass to power the radioactive tail of the light curves.\\
(4) The plateau phase of the light curves shows $\sim 0.53$~mag per 100 days decay rate in $V$ band, which is smaller that what is expected from the known decay rate--peak luminosity correlation \citep{Anderson2014}. This may indicate the early contribution of  $^{56}$Ni radioactive decay to the luminosity as discussed by \cite{Nakar2016}.\\
(5) From modelling of the light curves, the best-fit progenitor is a $\sim$18.8~M$_\sun$ ZAMS star that evolved into mass of $\sim 15$~M$_\sun$ and radius of $\sim 1040$~R$_\sun$ before an explosion of $\sim 1.3$~foe.\\
(6) The \kspn\ appears to have subsolar metallicity with  $\rm Z/Z_\Sun \simeq0.1-0.4$.\\
(7) Comparison between the observed light curves and modelled light curves of \kspn\ indicate that a CSM component may have been present around its progenitor.\\
(8) The production of $0.10\pm0.01$~M$_\sun$ of $^{56}$Ni mass in an explosion of $\sim 1.3$~foe is consistent with the predictions of the neutrino-deriven CCSN simulations of \cite{Sukhbold2016}.\\
(9) The estimated ZAMS mass of $\sim 18.8$~\Msolar for \kspn\ progenitor is higher than the range of \Mzams$\lesssim 17$~\Msolar\ obtained for the RSG progenitor of observed Type II SNe \citep[see e.g.,][for the RSG problem]{Smartt2009}. \\
(10) The large ZAMS mass of \kspn\ together with its subsolar metallicity is suggestive of a potential link between low metallicity progenitors and SN in massive RSGs \citep{Anderson2018}.

This research has made use of the KMTNet facility operated
by the Korea Astronomy and Space Science Institute
and the data were obtained at three host sites of CTIO in Chile,
SAAO in South Africa, and SSO in Australia. We acknowledge with thanks the variable star observations from the AAVSO International Database contributed by observers worldwide and used in this research. Our simulations were carried out on Compute Canada resources. PyRAF is product of the Space Telescope Science Institute, which is operated by AURA for NASA. N.A. was supported by  QEII-GSST and OGS Fellowships. D.S.M. was
supported in part by a Leading Edge Fund from the
Canadian Foundation for Innovation (project No. 30951) and a
Discovery Grant from the Natural Sciences and Engineering
Research Council of Canada. Support for this work was
provided to M.R.D. by NASA through Hubble Fellowship
grant NSG-HF2-51373 awarded by the Space Telescope
Science Institute, which is operated by the Association of
Universities for Research in Astronomy, Inc., for NASA, under
contract NAS5-26555. M.R.D. acknowledges support from the
Dunlap Institute at the University of Toronto.
SGG acknowledges the support from the Portuguese Strategic Programme 
UID/FIS/00099/2013 for CENTRA and the FCT project PTDC/FIS-AST/31546/2017.
Support for G.P. is provided by the Ministry of Economy, Development, and Tourism's Millennium Science Initiative through grant IC120009, awarded to The Millennium Institute of Astrophysics, MAS. Finally, we thank the anonymous referee for helpful comments on the original draft of this work.

\software{SNEC \citep{Morozova2015}, NumPy \citep{vanderWalt2011}, Matplotlib \citep{Hunter2007}, Astropy \citep{AstropyCollaboration2013,AstropyCollaboration2018}, FAST \citep{Kriek2009}, Hotpants \citep{Becker2015}, IRAF \citep{Tody1993}, SCAMP \citep{Bertin2006}, SWARP \citep{Bertin2002}, PyMCZ \citep{Bianco2016}, KEPLER \citep{Woosley2007}, ds9 \citep{Joye2003}, PyRAF}

\facility{KMTNet, DuPont, SciNet, AAVSO}

\pagebreak

\end{document}